**Review article**

# Circularity in Perovskite-Based Tandem Photovoltaics for Terawatt-Scale Deployment

Abderrahime Sekkat[1,2,*], Shiling Dong[2], Jenny Baker[3], Matt Burnell[4], Tapas Mallick[4,5], Ruy S. Bonilla[6], Robert L. Z. Hoye[2,*]

[1]University of Toulouse, Toulouse INP, CNRS, LGC, Toulouse, France

[2]Inorganic Chemistry Laboratory, University of Oxford, Oxford, United Kingdom

[3]Department of Mechanical Engineering, University of Bath, Bath, United Kingdom

[4]Environment and Sustainability Institute, University of Exeter, United Kingdom

[5]Department of Mechanical and Energy Engineering, College of Engineering, Imam Abdulrahman Bin Faisal University, Saudi Arabia.

[6]Department of Materials, University of Oxford, Oxford, OX1 3PH, United Kingdom

* E-mail: abderrahime.sekkat@toulouse-inp.fr (A.S.), robert.hoye@chem.ox.ac.uk (R.L.Z.H.)

## Abstract

As photovoltaics (PVs) scale from one to multiple terawatts over the next decade, ensuring sustainable deployment is urgently required. Crystalline silicon (c-Si) PVs, the current industry standard, will generate an estimated 160 million tonnes of waste by 2050, and there remains complex technoeconomic challenges associated with their recycling. Metal-halide perovskite (MHP)–based tandem PVs not only promise higher power conversion efficiencies than single-junction c-Si devices, but also offer intrinsic advantages for circularity, including simpler device architectures, low-temperature processing, and more accessible materials recovery routes. At this pivotal juncture when perovskite PVs begin to enter the market, this review examines the critical circularity challenges that must be addressed: substitution of scarce raw materials, scalable recycling protocols, cost-effective stack delamination, safe lead sequestration, and policy frameworks to encourage circularity across the device lifecycle with effective incentives. By integrating the materials, technoeconomic and policy dimensions that go beyond conventional lifecycle assessments, we outline actionable strategies to co-optimize device performance and sustainability. This review aims to guide researchers, policymakers, and industry stakeholders in steering perovskite-based tandem PVs towards a circular and responsible commercialization pathway within the global clean-energy transition.

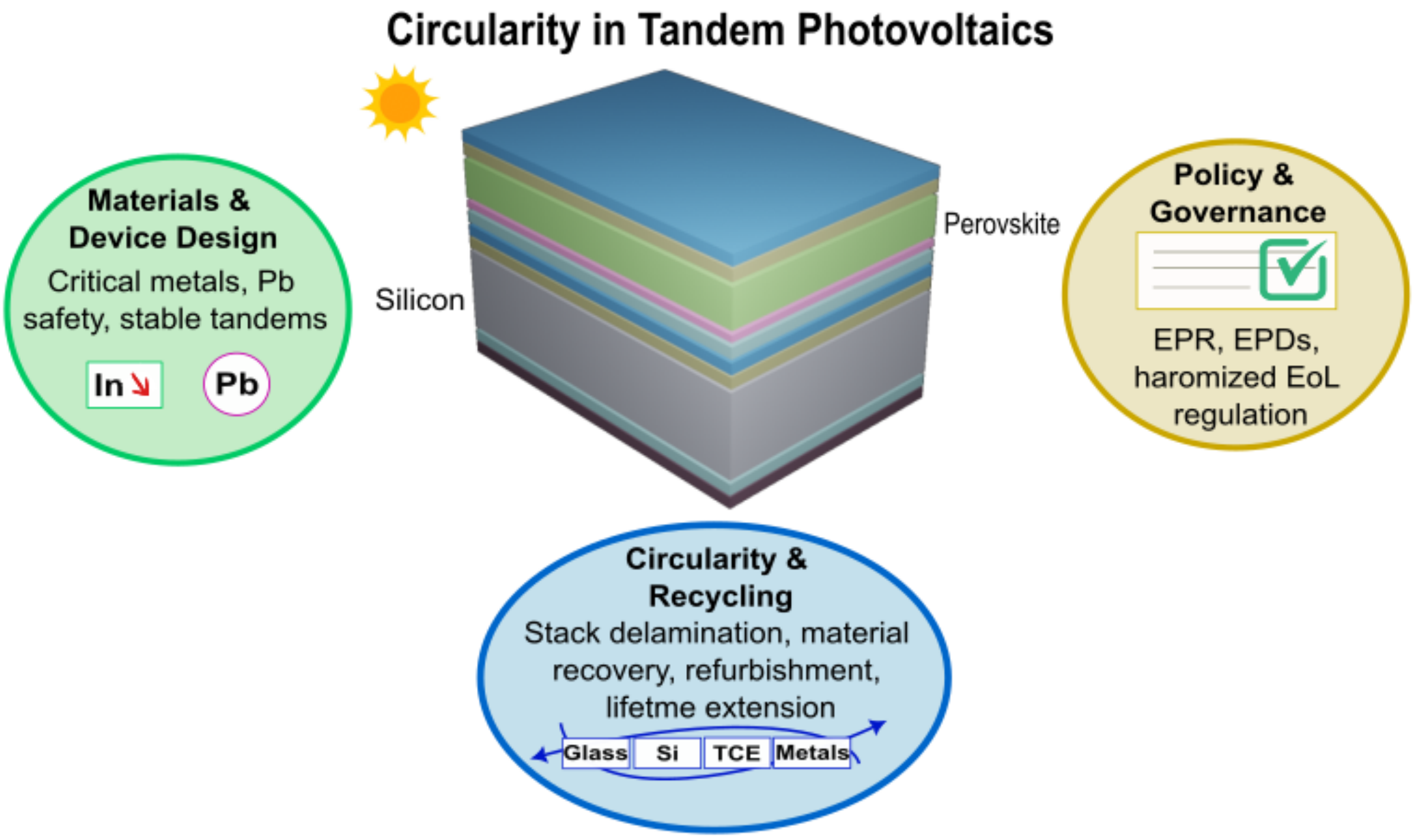

## [H1] Introduction

The global deployment of photovoltaic (PV) technologies has entered an unprecedented phase of rapid expansion, with cumulative capacities expected to produce multiple terawatts of power ($TW_p$) within the next decade (**Figure 1a**)[1,2]. This scale-up is critical to fulfilling net-zero climate targets across the world[3–5]. But with this rapid growth, there are urgent sustainability challenges, particularly regarding the management of PV waste and the circularity of materials used in solar modules (**Figure 1b**). Currently, industry-dominant crystalline silicon (c-Si) PVs face significant limitations in recycling[6–8]. Despite silicon and precious metals like silver (Ag) being valuable resources, only 15–20% of these materials are recovered from end-of-life (EOL) modules[9]. The complex construction of c-Si panels, including durable encapsulants and multi-layered assemblies, presents substantial technical and economic barriers to effective recycling (**Figure 1c**). As a result, many panels in the USA are disposed of via landfill, while in Europe they are often repurposed, for instance, as building aggregates in roads, in ways that limit true materials circularity and continue to pose environmental challenges[8,10,11]. Although glass accounts for more than 70% of module mass and can be recovered at rates above 96%[12–14], due to contamination, this glass is typically downgraded to cement or aggregates with reduced environmental benefits compared to recycling glass back to sheet[15]. In addition, the depletion of critical resources (especially silver, indium, copper), energy- and carbon-intensive production, and EOL waste management, remain unaddressed for multi-$TW_p$-scale PV deployment[16].

In this context, metal-halide perovskite (MHP)–based tandem solar cells have emerged as a promising technology with the potential to complement existing c-Si PV systems. Multi-junction devices combine sub-cells absorbing in complementary parts of the solar spectrum, raising theoretical efficiency limits well beyond the practical limit of single-junction c-Si PVs (considered to be 28%)[17–19]. In this case, Pb-halide perovskites absorb in the visible part of the solar spectrum, while c-Si absorbs in the near-infrared part of the solar spectrum. Laboratory-scale demonstration of perovskite/Si tandems with 1 $cm^2$ device area have already surpassed 34% efficiency[20], with pathways toward 37% and beyond[21], offering a route to higher power generation per unit area, and lower levelized cost of energy (LCOE) than single-junction c-Si PVs. These efficiency gains translate into higher energy yield per unit area, which therefore lowers the environmental impact per kWh generated. Furthermore, the c-Si bottom cell can be replaced with a thin film near-infrared absorber (*e.g.*, Pb/Sn alloy perovskites, such as $(Cs,FA,MA)Sn_{0.5}Pb_{0.3}I_3$, which has a bandgap of 1.26 eV[22]). Replacing the c-Si cell can allow MHP-based tandems to overcome the challenges inherent to c-Si modules, such as its high carbon dioxide equivalent ($CO_2$eq) footprint, high energy cost of production, and challenging and costly recycling. In particular, perovskite cells require lower material use than c-Si cells, and can be processed via low-temperature methods that enable reduced $CO_2$eq footprints during manufacture[23]. Furthermore, the functional layers in MHP cells are largely soluble in relatively mild solvents, which opens up recycling routes for the MHP cell(s)[23,24] that are potentially simpler and more cost-effective than c-Si recycling[25,26].

These recycling processes possible with MHP-based PVs could enable closed-loop material recovery and reduce the environmental footprint of solar energy systems[27,28]. As a result, the MHP cell(s) in perovskite/Si tandems could be removed at EOL from the tandem, and the longer-lived c-Si cell reused; alternatively, all-perovskite thin film tandems could be entirely dissolved at EOL, potentially overcoming challenges with waste accumulation from c-Si modules. Nevertheless, realizing these benefits and demonstrations made in the lab to commercial advantages requires addressing several key factors[16]. As MHP-based tandem technologies are still at an early stage, now is the pivotal time to shape their future deployment by balancing lifecycle considerations with efficiency improvements. These concerns include (**i**) the practicalities of separating modules to access the active materials for recycling, and whether the devices/materials extracted are of sufficient value for re-use in commercially-competitive modules, **(ii)** the use of critical elements in the device architecture, **(iii)** preventing harmful release of Pb-containing compounds to the environment during manufacturing, deployment and EOL, **(iv)** reducing the energy intensity and environmental impacts of the manufacturing processes, **(v)** the processing sequences required for recycling, and (**vi**) policy frameworks ensuring sustainable EOL management, driving mandates for high recovery rates, and fostering resilient supply chains.

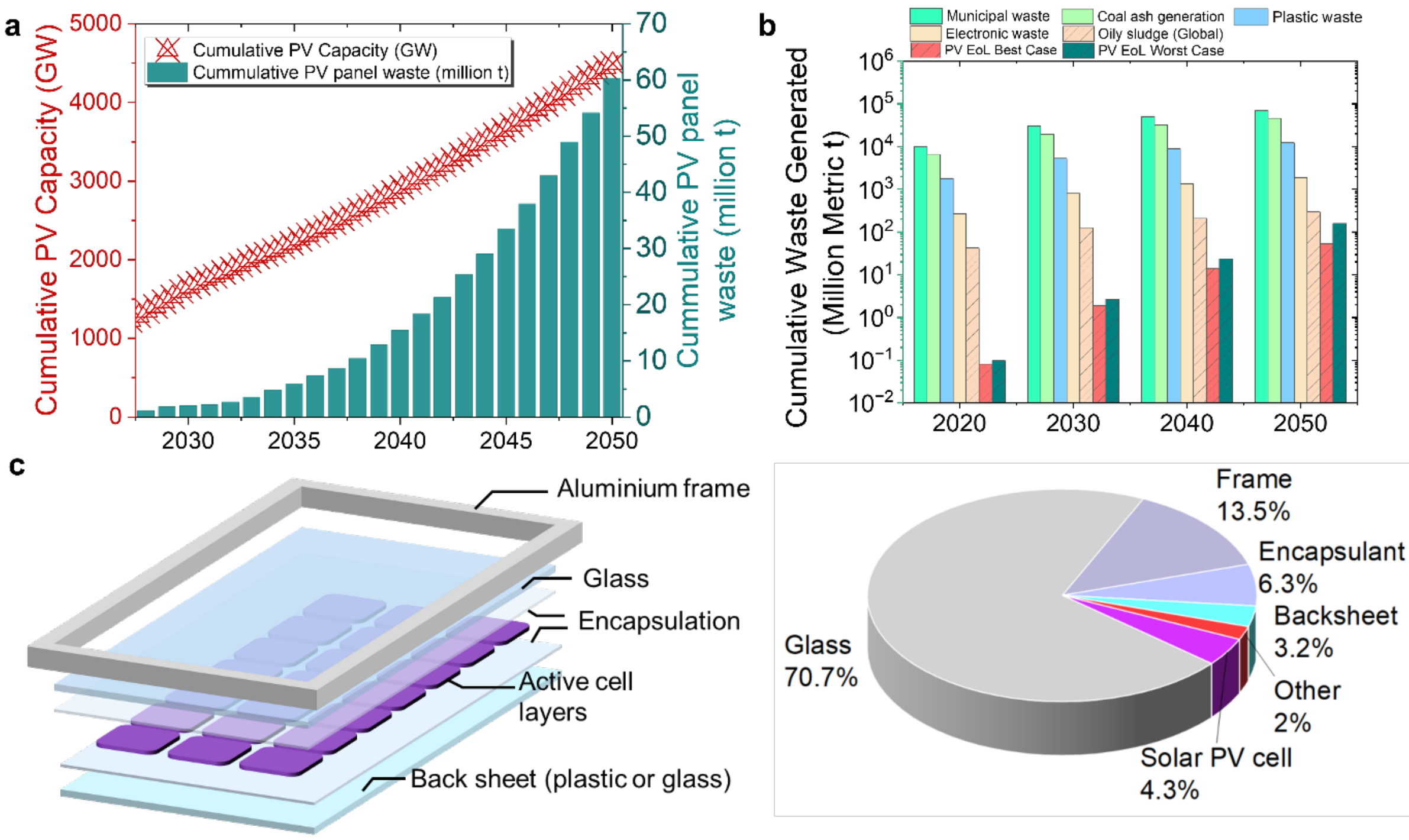

**Fig. 1 | Waste from terawatt-scale PV deployment. a,** Estimated cumulative global waste volumes of end-of-life PV modules and PV capacity. Adapted with permission from Ref. 29. **b,** Global cumulative waste produced from 2016 to 2050, including municipal waste, coal ash, plastic waster, e-waster, oily sludge, and best-/worst-case scenario PV water estimates from modules, and **c,** diagrammatic representation of the structure composition of a silicon solar panel and solar cell. The typical weight fraction of each component in the c-Si module is suggested. Adapted with permission from Ref. 30.

This review aims to comprehensively discuss the potential and challenges of MHP-based tandem photovoltaics from a circularity perspective, drawing together materials science, technoeconomic, and policy factors. The primary focus is on MHP/Si tandems, since these are closest to commercialization, but we also cover emerging MHP/MHP tandems. We begin by holistically discussing the circularity and sustainability challenges of the current PV industry, and the limitations of the current policy framework. Subsequently, we discuss the opportunities opened up by perovskite-based tandems for improved circularity, but also the materials challenges introduced, and strategies to overcome these. Moving beyond classical lifecycle assessments (LCA) goals[3,16,20–2], we systematically evaluate how tandem PV technologies can be developed and deployed in a manner that aligns with broader sustainability goals, including minimizing lifecycle environmental burdens, reducing dependence on scarce or toxic materials, and ensuring responsible EOL management[5,31–33]. In contrast to previous reviews that focus primarily on manufacturing sustainability or single-junction perovskites[30,34–41], we provide an integrated assessment of circularity strategies tailored specifically to tandem PVs, detailing practical pathways for recycling MHP/Si tandems, exploring the largely unaddressed opportunities and challenges of recycling MHP/MHP tandems, and evaluating how these approaches interact with materials criticality and policy constraints[28,38,42].

## [H1] Sustainability Challenges in the Photovoltaics Industry

In this section, we begin the review by discussing the sustainability challenges in the c-Si PV supply chain, which will remain in place for perovskite/c-Si tandems, and have to be addressed. A critical challenge is to reduce the embodied $CO_2$eq footprint across the entire lifecycle[43], including the mining and extraction of virgin materials used in c-Si production, and the use of critical raw materials (CRMs), such as Si, Ag and Cu[44,45]. Harmful or regulated elements (Pb, As, Sb), as described by the British mandatory classification and labelling list[46], are found in some components of c-Si modules, *e.g.*, the Pb/Sn solder used in wiring[47] and Sb used in solar glass. MHP-based PVs also contain critical elements, including In and Cs (in addition to Pb, Sb), which can impose significant supply-risk, embodied environmental impact, and the risk of ecological damage if not handled properly at EOL[27,43]. The manufacturing of c-Si panels has largely been 'offshored' to Asia via a centralized model[43,48], contributing to $CO_2$eq emissions associated with production in regions where electricity supply is via high $CO_2$eq sources[27,28]. The transportation of solar modules to global markets contributes to the overall emissions of the solar value chain[49] with international shipping accounting for approximately 2.2–2.1% of total global greenhouse gas (GHG) emissions on a $CO_2$eq basis[50]. Transitioning to MHP-based tandems increases the power-to-weight and power-to-area ratio due to the higher energy yield offered by tandem PVs, which in turn reduces the overall impact from transportation from manufacturer to end-user compared to single-junction c-Si PVs. For MHP-based tandems to have improved sustainability credentials compared to current c-Si modules, tandem manufacturing should take place in regions close to deployment, using electricity supplies with relatively low $CO_2$eq sources, and using locally-sourced materials with harmful materials either eliminated or heavily managed at EOL. Although Environmental Product Declarations and regulatory trends (especially in the EU) are moving toward mandatory lifecycle impact disclosure, publishing detailed comparative and uniform LCA data on c-Si PV datasheets is rare and not currently required worldwide, and without intervention, MHP-based tandems are on the same policy trajectory. Upon widescale deployment of these tandems, under the existing global PV framework, consumers would not be able to objectively compare environmental credentials of c-Si with MHP-based technologies.

The inevitably high volumes of waste from c-Si modules remains a question of timing with regard to when decommissioning will occur at a large-scale[29,51,52]. The longevity of c-Si modules is expected to be 25-30 years, largely because of decades-long optimization of encapsulation methods[53], which prolongs the time taken before waste is produced compared to other Electrical and Electronic Equipment (EEE)[54]. In comparison, the longevity of MHPs is less certain due to the nascency of this technology, creating an inherent sustainability challenge that the MHP cell may reach EOL years before the c-Si cell. Premature EOL of the MHP sub-cell could negate the overall sustainability benefits gained from the transition to tandems with MHPs, and have the potential to exacerbate exponential c-Si waste volumes expected during this period, unless effective methods for recycling the c-Si bottom cell are developed.

Furthermore, an unintended consequence of transitioning from single-junction c-Si PV to MHP-based tandems is that older, working c-Si systems are prematurely decommissioned and replaced with newer tandem modules, further accelerating waste volumes unless viable c-Si reuse markets can be established. Paradoxically, the deployment of these higher-efficiency next-generation MHP-based tandem technologies could drive premature decommissioning of legacy single-junction c-Si PV systems, as electricity end-users, particularly commercial consumers, replace existing installations to increase on-site energy yields[51]. This is driven by several factors, especially the increasing efficiency of new panels, and the reduced cost of PV compared with auxiliary infrastructure, such as frames, grid connection, and inverters[55]. The environmental impacts of repowering using MHP/Si tandems shows that whilst in some scenarios a financial benefit is obtained by replacing c-Si single-junction modules after 8 years of service, a $CO_{2eq}$ emissions benefit (on replacement with higher-efficiency panels) is not typically realized until 15 years of service[55]. For the greatest impact on reducing global emissions, high-efficiency PV devices, such as MHP/Si tandems, should be deployed to maximize fossil fuels displacement, for example, in countries with high coal use, such as Poland, Cyprus or Greece[55]. Without policy intervention, MHP-based tandems are likely to shorten the lifetime of current single junction PV in countries with strong incentives for low carbon electricity generation and lower carbon emissions[55]. In this respect, keeping c-Si modules operational for as long as possible reduces the overall embodied $CO_2$eq footprint of the primary c-Si and secondary MHP supply chain. Policy must drive incentives for MHP-based tandems to be preferentially used as new installations, or penalizing premature c-Si single-junction PV decommissioning. Within this model, the waste hierarchy is observed, with recycling only considered an act of 'last resort'.

When the c-Si and MHP-based tandem modules eventually reach EOL, the energy intensity needed for 'clean' delamination and separation of the module into its constituent parts[56,57] is currently not well quantified. The transportation of bulky, heavy c-Si and MHP tandem panels to the centralized PV recycling infrastructure also adds a hidden $CO_2$eq cost to the total emissions[58,59], although less-so for MHP-based tandems due to their higher efficiencies. While this $CO_2$eq contribution is relatively small compared to total lifecycle emissions, 'first-mile' transport costs to central distribution centres can be high, particularly in remote regions.

Under the existing Extended Producer Responsibility (EPR) framework, PV producers are not incentivized to design for circularity[60], leaving the burden on future generations to develop economically feasible recycling solutions. There is no mechanism for manufacturers to be held accountable today for the products that will enter the waste stream in over 25 years time, by which time the manufacturer may cease to exist. As per Box 1, decision makers have various policy instruments to invigorate the circular economy and sustainability throughout the solar product lifecycle. However, the timeframe for policy to be implemented and international standards to be recognized that would alleviate sustainability challenges for PV technologies is a concern compared to the speed at which the technology itself is advancing.

To summarize, for future solar technologies to be sustainable, consideration must be given for the responsible sourcing of materials and location of manufacturing in relation to deployment, with traceability and transparency integral for the information provided to consumers. Manufacturers must be incentivized (or regulated) to produce easier-to-recycle technologies, whilst not compromising the longevity or performance, and with recycling undertaken as close to the source of the waste as possible, and a policy framework which preempts and embraces the technological advancements.

**Box 1 | Policy Frameworks for Circularity**

Overarching circular economy (CE) principles should aim to eliminate waste and pollution, keep products and materials in use, and regenerate natural systems[61]. CE policy instruments can be reduced down to five categories[62]: i) *regulatory* instruments are those that set mandatory practices or targets; ii) *economic* instruments are those that set financial incentives or disincentives; iii) *technical* instruments establish technical standards on how to achieve circularity, usually through design; iv) *operational* instruments are those that create facilitating infrastructure or processes; v) *communicative* – focusing on education and knowledge dissemination. Policy falls within six general categories of the PV lifecycle (figure below) and can also be included with the scope of wider aspirational policy frameworks such as the EU Green Deal and the Circular Economy Action Plan[63].

1. Design policy considers aspects such as performance, utilisation, recyclability, repairability, destructibility, durability, material inputs, and material integrity. Design must conform to existing international standards on PV product safety and product viability.
2. Material input policy includes the traceability of environmental and social impacts of extracting virgin/recycled materials and subsequent value chain to the PV manufacturing facility.
3. The production policy stage considers reporting and minimizing the overall energy usage (*e.g.*, embodied $CO_2$eq), toxicity of embedded materials and waste generated during the manufacturing process, and subsequent global distribution of PV modules. Extended Producer Responsibility is enforced at this stage.
4. Use and consumption policy concerns presenting information to the end-user to make informed decisions and extending the lifetime of the product to prevent waste from occurring in the first place.
5. Resource circulation policy considers the overarching definitions of waste. At this stage reuse and repair are considered, along with liability and financing of waste processes.
6. Waste policy concerns the correct protocols once PV material is defined as waste, such as transportation, reporting, recycling methodology and environmental protection.

Policy is limited by variations in geographic jurisdiction and the conflict between the locality of the manufacturer and the global market. Buyers cannot seamlessly specify to manufacturers the environmental credentials of the PV technology they require, and the speed at which policy is developed and ratified rarely keeps pace with technological advancements. There is also reticence from policymakers to use economic instruments, as this inevitably raises the cost of PV technologies, which conflicts with the overarching aim of increased renewable deployment and energy independence as a wider geopolitical strategy.

In the EU, the major policy instrument that is used for c-Si recycling schemes comes under EPR legislation, such as the WEEE Regulations. EPR mandates manufacturers to provide information about the potential environmental impacts of their products, to be responsible for the physical handling, and for covering the cost associated with end-of-life management of their products[64]. However, EPR relies on enforcement by the authorities to be properly enacted, and there is evidence that PV producers are 'freeriding'[64] their responsibilities to finance EPR[65].

EPR does not mandate adequate eco-design strategies ('upstream'), instead placing the burden of creating technoeconomic solutions for recycling at end-of-life ('downstream') on future generations to resolve. EPR does not mandate PV manufacturers to release information of embedded materials or recycling methods. Waste regulation is typically slow to be enforced in relation to the pace of technological change, rendering it outdated by the time it is implemented.

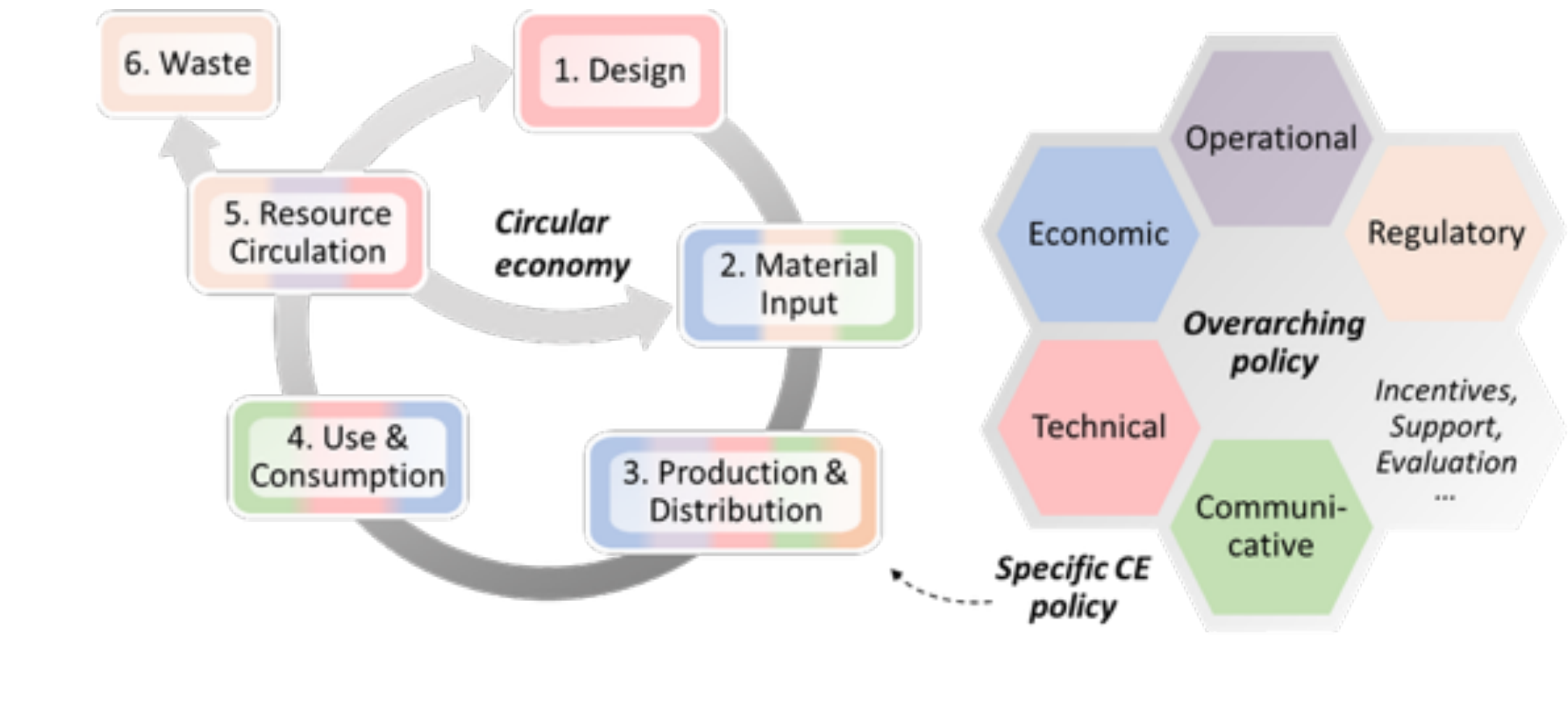

## [H1] Opportunities with Perovskite-Based Tandems

MHPs have emerged as promising solar absorbers thanks to their bandgap tuneability, low temperature processability and cost-effectiveness[66]. These properties make perovskites attractive both as single-junction photovoltaic absorbers and as tunable top cells in tandem architectures, where their bandgap can be optimized to complement a lower-bandgap bottom cell. Their versatility supports their integration into diverse tandem configurations, including with c-Si[67], copper indium gallium diselenide (CIGSe)[68], kesterites[69], cadmium telluride (CdTe)[70], organic PV[71], and all-perovskite combinations[22] (**Box 2**). By mid-2025, perovskite/Si two-terminal tandem solar cells have achieved record efficiencies of 34.85% (modules up to 33%), surpassing the efficiency limits of single-junction c-Si PVs[72]. All-perovskite tandems have reached 30.1% power conversion efficiency (PCE)[72], and have a realistic path toward >37% efficiency and commercial viability[22,73]. Due to their increasing commercial relevance and potential for deployment at the level of >1 $TW_p$ $yr^{-1}$, this review focuses on double-junction MHP/Si tandems and all-MHP tandem PVs. Such MHP-based tandem technologies combine the high efficiency potential of tandem architectures with the low-temperature processing and thin-film material use characteristic of perovskite absorbers, creating opportunities to meet performance requirements while delivering sustainability improvements. We examine the prospective potential for reductions in primary manufacturing energy demand, material intensity, and improvements in recyclability compared with single-junction c-Si solar cells.

**Box 2 | Device Architectures and Raw Materials in Perovskite-Based Tandems**

Tandem solar cells come in a variety of architectures. They are first classified by the number of sub-cells (commonly referred to as junctions). Two junctions absorb light from two separate regions of the solar spectrum, halving the current but increasing the voltage by 2.3–2.6 times, hence a gain in total power output to a theoretical maximum of 45%[74]. More junctions are possible by using 3 or more absorbers, but gains in efficiency start to decline[74] (3 junctions ~50%, 4 junctions ~54%).

The second classifier is the number of contacts: 2-, 3- or 4-terminal refer to the number of electrodes that connect the tandem device to the external circuit for current extraction and power generation. Four terminal (4-T) tandems are assembled via 2 independent sub-cells, each with its own wiring. In three terminal (3-T) tandems, the two sub-cells share a common electrode that can balance the current generated on the two cells. 2-T tandems are stacked in series and interconnected via a transparent conducting electrode (TCE), such that the same current flows through both sub-cells. The overall tandem cell current is given by the sub-cell that generates the smaller current, and hence in 2-T tandems, the two sub-cells need careful current matching. The 2-T device architecture comes with the advantages of fewer external wires, ease of integration with standard power electronics, and savings both in efficiency and material use from reducing the use of TCEs. Due to their relevance and close-to-market potential, we focus on 2-T/2-junction tandem solar cells.

The third classification is given by the solar absorber material used to assemble the two sub-cells. In 2-J, the top layer contains a wide-bandgap (WBG) semiconductor that absorbs higher-energy photons, while the bottom layer uses a narrow-bandgap (NBG) material to capture lower-energy photons. The tandem can therefore be made with all-perovskite junctions, a WBG perovskite with NBG c-Si, a WBG perovskite combined with another NBG thin film PV such as organic solar cells or CdTe PV. Among these possible combinations, MHP/Si tandems are the closest to commercial use, with substantial investments being made worldwide to manufacture this technology at scale[75–77].

The WBG perovskite is comprised of a monovalent cation in the A-site (typically formamidinium ($FA^+$), cesium ($Cs^+$), or a combination of these two), and iodide ($I^-$) and bromide ($Br^-$) in the X site to reach 1.6–1.7 eV band gap. Electron- and hole-transport layers (ETL/HTL) are also required at each side of the cell to enable preferential extraction of one charge-carrier. These charge-transport layers are often so thin that material usage is negligible in comparison to the rest of the device. The lead halide perovskite (LHP) device requires at least one TCE, which are currently most commonly comprised of materials based on indium oxide ($In_2O_3$). The NBG silicon sub-cell can take the form of well-known silicon heterojunction (SHJ) or tunnelling oxide passivating contact (TOPCon) architectures. These might require one or two TCEs, but the main material in (mg $W^{-1}$) remains the 150–200 µm thick Si wafer.

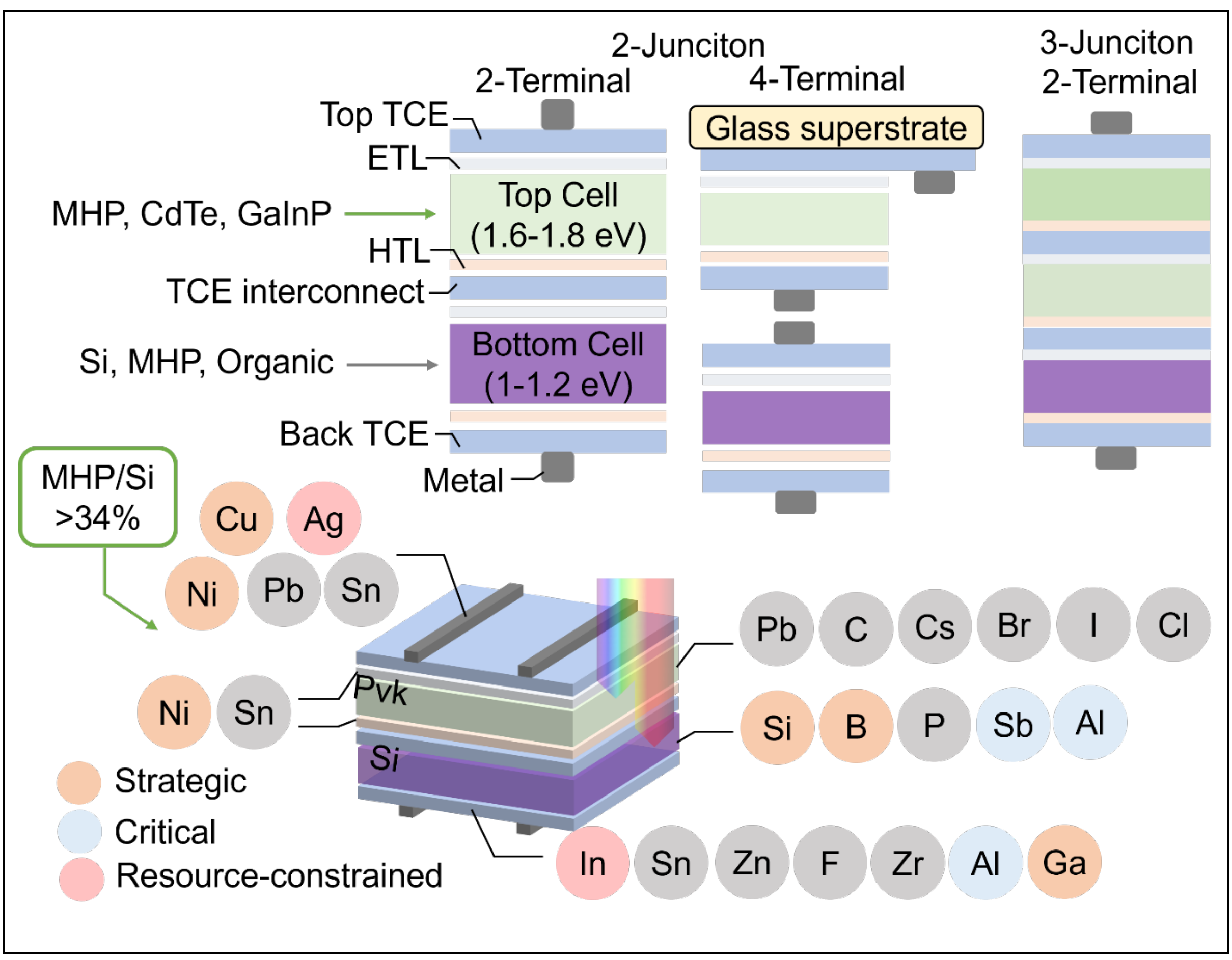

*[H2] Improved Energy Yield*

Tandem solar cells have the potential to increase PCE from ~25–26% (c-Si single junction PV) to a theoretical 45% PCE for two junctions, and >50% for three junctions[78,79], drastically improving electricity production and optimizing land use[80]. Optimized MHP/Si PV systems can achieve annual energy yields (electrical energy produced by PVs, in kWh, in real-world scenarios)[81] up to 30% higher than single-junction c-Si PVs under real-world conditions[82]. The energy yields from PVs are influenced by geographic location, solar irradiance, spectral distribution, operating temperatures and aerosol within the atmosphere. In contrast to single junction c-Si cells, MHP/Si devices allow the optimization of energy yield based on the direction, intensity, and spectrum of solar radiation onsite. Additionally, temperature fluctuations impact the open-circuit voltage and efficiency of both silicon and perovskite cells, but perovskites exhibit a lower temperature coefficient (typically –0.15% to –0.3% $°C^{-1}$ vs. silicon's –0.4% to –0.5% $°C^{-1}$), providing a relative advantage for perovskite-based devices in warmer conditions[83–86]. By tailoring the design of perovskite solar cells, primarily through bandgap tuning (e.g., via compositional changes) and absorber layer thickness optimization[87], to specific locations and operating conditions, it is possible to maximize their energy output and achieve improved material utilization ratios (in mg $kWh^{-1}$).

MHP/Si and MHP/MHP tandem PVs also offer advantages for rooftop and urban applications where space is limited[78]. Additionally, perovskite cells are lighter than their c-Si counterparts, and all-thin film tandems thus enable low-weight, high-specific-power modules manufactured via low-temperature processes that facilitate easier integration into existing building structures and reduce installation costs. While utility-scale applications benefit from added energy yield alongside economies of scale to minimize LCOE, residential and commercial rooftops prioritize efficiency per unit area and seamless integration, areas where perovskite-based tandems, especially all-thin film designs, excel by maximizing output from constrained spaces.

Altogether, the higher energy yield in MHP-based tandem PVs translates directly into sustainability benefits by delivering more electricity per manufacturing energy and material input than conventional c-Si PVs, thereby reducing balance-of-system requirements, land use, and lifecycle greenhouse gas emissions per unit power generated. Owing to the low temperature processability of the MHP top-cell, perovskite-silicon tandems can achieve shorter energy payback times and lower embodied energy than single-junction c-Si PVs[88]. This combination of enhanced energy yield and reduced material and manufacturing intensity makes perovskite-based tandems especially promising for high-performance PV deployment than c-Si alone.

### *[H2] Reduced Environmental Impact*

The materials and processing of perovskite-based tandems show key challenges and opportunities in relation to sustainability. Most high-efficiency MHP-based tandems use lead-based absorbers and indium-based TCEs, raising environmental and supply-chain concerns[89]. Although work on lead-free perovskites continues, these alternatives currently lag behind in efficiency and (for Sn-based perovskites) stability compared to Pb-halide perovskite systems[89–92], and do not necessarily have lower environmental impacts as a result of these lower energy yields and device lifetime[93]. At the same time, the thin-film nature and low-temperature processing of perovskites offer intrinsic advantages in reducing material use and embodied energy relative to wafer-based photovoltaics. Lab-scale (spin coated) solution-based manufacturing of MHP-based tandems does not scale to gigawatt-level production[94], and so alternative methods are under development to achieve uniform, defect-free film deposition over large-area, textured silicon wafer bottom cells[94–96], such as through thermal co-evaporation, a combination of evaporation and solution processing, and scalable meniscus-coating techniques like slot-die coating[97,98]. However, production throughput, yield, and reproducibility continue to limit scalability, which in turn affects device lifetime and energy output, and therefore critically influence overall life-cycle environmental performance[94].

On the other hand, early life-cycle assessments already suggest that these processing advantages could offer reduced embodied energy and faster payback times than single-junction c-Si PV when combined with improved device stability, precursor utilization efficiency, and manufacturing yield[99]. Embedded energy and greenhouse gas (GHG) footprint metrics for MHP-based tandem PV show their competitiveness over c-Si PV devices (Box 3). Based on a survey of previously published LCA models (summarized in Supplementary Note 1; full assumptions of insolation, electricity mix, system boundary, lifetime, performance ratio detailed in Supplementary Data), the median primary energy demand for a laboratory-scale MHP/Si tandem module is 13 840 MJ $kW_p^{-1}$, with the dominant contributions to primary energy demand arising from wafer-based silicon and module glass components[88]. Under a representative insolation of 1700 kWh $m^{-2}$ $yr^{-1}$, assuming a performance ratio of 0.8 and a standard grid-efficiency correction, this energy demand corresponds to a median energy-payback time (EPBT) of 1.3 years and global warming potential (GWP) of 915 $gCO_2eq$ $kW_p^{-1}$. These are still marginally improved over c-Si single-junction PVs, which have typical values of 1.5 years and 1224 g $CO_2eq$ $kW_p^{-1}$, respectively. But switching to all-perovskite tandems substantially reduces both indicators, with median a EPBT of 0.33 years and GHG intensity of 209 $gCO_2eq$ $kW_p^{-1}$ [88,100]. These improvements arise from thin printable absorbers, lower-temperature fabrication for the entire tandem device, and the elimination of energy-intensive silicon purification. Other reviews have highlight three dominant levers for further progress: solvent recovery, roll-to-roll deposition, and lifetime extension[5]. Extending service life from 10 to 25 years halves GWP while unalters EPBT[88,101].

Lead remains an environmental issue, yet the mass contained with a tandem module is below 0.003 g $Wp^{-1}$. This low material inventory, combined with encapsulation, sequestration and existing industrial recycling infrastructures, makes closed-loop recovery technically feasible. Industrial pilot lines are validating these environmental assessments, and outdoor prototypes already pass standard damp-heat and thermal-cycling industrial tests[16]. MHP-based tandems are still at an early commercial stage but shows great promise in overcoming these technical and manufacturing challenges[72,99,102].

### *[H2] Recycling Strategies*

The primary economic opportunity in recycling MHP-based tandem modules lies in the recovery of the c-Si bottom cell (MHP/Si tandems) and indium tin oxide (ITO) substrates (all-thin film tandems), which represent the largest share of the device's environmental impact[103]. 4-T tandems offer recycling advantage because each sub-cell is fabricated independently and mechanically stacked, allowing for easy separation at EOL of either sub-cell.

However, 2-T tandems are currently preferred in efforts to commercialize perovskite/Si tandems because of lower requirements for the transparent conducting electrodes, reduced optical losses (and therefore higher efficiency limits), and do not require additional inverters, thus lowering overall balance of system costs[92,104,105]. Recovering various highly valuable materials from 2-T tandem solar cells is significantly more complicated due to the monolithic integration of the sub-cells. Thermal and mechanical delamination strategies allow for the physical separation of Si from the MHP top layer while retaining the optoelectronic quality of the c-Si layer. A two-step protocol was reported recovering the c-Si bottom cell from a degraded MHP/Si tandem: thermal delamination at approximately 180 °C to soften the encapsulant and mechanically separating the top MHP sub-cell from the bottom c-Si wafer, followed by selective chemical cleaning to strip residual interlayers[106]. The tandems built on these recovered c-Si cells show a PCE of 25.7%, which is comparable to those fabricated from the fresh cells (26.3%)[106]. Multi-solvent chemical baths can selectively dissolve the functional layers. HTL materials, *e.g.*, 2,2',7,7'-tetrakis(N,N-di-p-methoxyphenylamine)-9,9'-spirobifluorene (spiro-OMeTAD), soluble in non-polar solvents, have been recycled using chlorobenzene, although dopants such as cobalt-bis(trifluoromethanesulfonyl)imide (Co-TFSI) were not removed by the process[107]. The perovskite layers can be dissolved by polar organic solvents, such *N*,*N*-dimethylformamide (DMF) or methylamine, although the hazardous solvents pose environmental risks and exhibit limited selectivity across diverse device stacks. Water-based recycling methods using specific additives (sodium acetate, sodium iodide, and hypophosphorous acid) and temperature control have been reported to recover functional layers, including lead(II) iodide ($PbI_2$), gold (Au), Tin oxide ($SnO_2$), Spiro-OMeTAD, and the ITO-coated glass[23,24], which significantly reduce overall resource depletion and lower the LCOE.

Conversely, MHP/MHP tandems are generally considered to have similar recycling method as single cell MHP solar cell, where the focus is reusing ITO-coated glass, recovery of Pb as $PbI_2$, and recycling organic solvents[108]. The thin interconnection layer (ICL) is not considered for recycling due to the low mass percentage and lower chances of selective removal of individual MHP layer followed by cost-effective separation of high-purity critical materials. However, separating the $SnI_2$ and $PbI_2$ salts from all-perovskite tandems is challenging, given that both Pb-only and Pb/Sn perovskites dissolve in the same polar solvents. Closed-loop manufacturing of all-perovskite tandems has not yet been demonstrated because of this challenge, and requires consideration of approaches to physically separate these two salts.

**Box 3 | Life Cycle Impact of Photovoltaic Panels**

LCAs quantify the environmental impact of technologies relative to a functional unit. In the case of PVs, the most basic unit is environmental impact per $m^2$, but this does not account for performance, geometric fill factor, device lifetime or location. Therefore, calculating impact per unit area is typically used for emerging technologies to assess material impact, especially when future performance improvements are expected. For more mature technologies, the impact per peak power produced by the device (*i.e.*, watt-peak, or $W_p$) is more commonly used. This allows direct PV panel comparisons between location or lifespan. To go a step further and compare impact across electricity sources (*e.g.*, wind, gas or hydroelectric sources of electricity), the impact per unit energy produced (in kWh) is needed. For PV, this depends on solar insolation and panel lifetime. Solar insolation can vary by over 300% globally, for instance, the UK averages at 900 kWh $m^{-2}$ $y^{-1}$ global horizontal irradiance (GHI) whereas regions such as Chile can average 3000 kWh $m^{-2}$ $y^{-1}$ GHI[109]. Panel lifetime, on the other hand, is affected not only by the durability of the module, but also on the reliability of auxiliary systems, such as inverters or batteries. Thus, when comparing reports of environment impact per kWh, it is essential that these studies state the assumed lifetime and insolation values to ensure comparability.

To convert between these functional units, we can use the following equations:

Example calculation if an impact is given per $m^2$ to convert to $kW_p$

$$Environmental\ impact\ per\ kWp = \frac{Environnmental\ impact\ per\ m^2\ \times 100}{\%\ Cell\ efficiency\ \times\ \%\ cell\ active\ area}$$

$$Impact\ per\ kWh\ produced = \frac{Impact\ per\ kWp}{(insolation\ per\ annum\ \times lifetime\ \times \sum_{n=0}^{19}(1-d)^n)}$$

Degradation is usually accounted for using a linear degradation rate. $d$ is the annual degradation rate (*e.g.*, 0.005 for 0.5% per year). $n$ is year index (from 0 to 19 for 20 years) When referring to power output in kWp, this is the panel's power output under 1000 W/$m^2$ solar insolation under the AM 1.5G spectrum at 25 °C[110].

To put perovskite-based single-cell and tandems in context, we compare their environmental impact with other single-junction PV technologies. We screened the LCA results across the literature, published values were first summarized using box plots and then re-plotted as scatter points with asymmetric error bars, where the median represents the central tendency and the interquartile range reflects the spread of reported values (detailed analysis in Supplementary Note 1). Two key impact indicators are considered: carbon footprint, expressed as carbon dioxide equivalent ($CO_2$eq), and cumulative energy demand per $m^2$ and per Wp, respectively. As shown in the left part of the figure below, the cumulative energy demand of silicon-free perovskite tandems is nearly six times lower than that of Si-based tandems, a difference attributed almost entirely to the energy-intensive silicon wafer refinement cycle, which dominates the impact of the latter[111]. This also translate to the carbon footprint of Si-free tandem approximately 2.5 times smaller than Si-based tandem (see right part of figure below), where the main impacts stem from manufacturing energy (notably vacuum deposition steps), deposition of transparent conducting oxides, and encapsulation[112–115]. The life-cycle emissions also depend sensitively on the $CO_2$eq intensity of the electricity grid used. It is crucial to recognize that due to the low technology readiness level of MHP-based. tandems, current LCAs are preliminary and often biased towards high-PCE laboratory devices. Most assessments have narrow system boundaries, excluding the impacts of balance-of-system components (inverters, wiring), transportation, and end-of-life processing, which are all vital for a complete environmental profile.

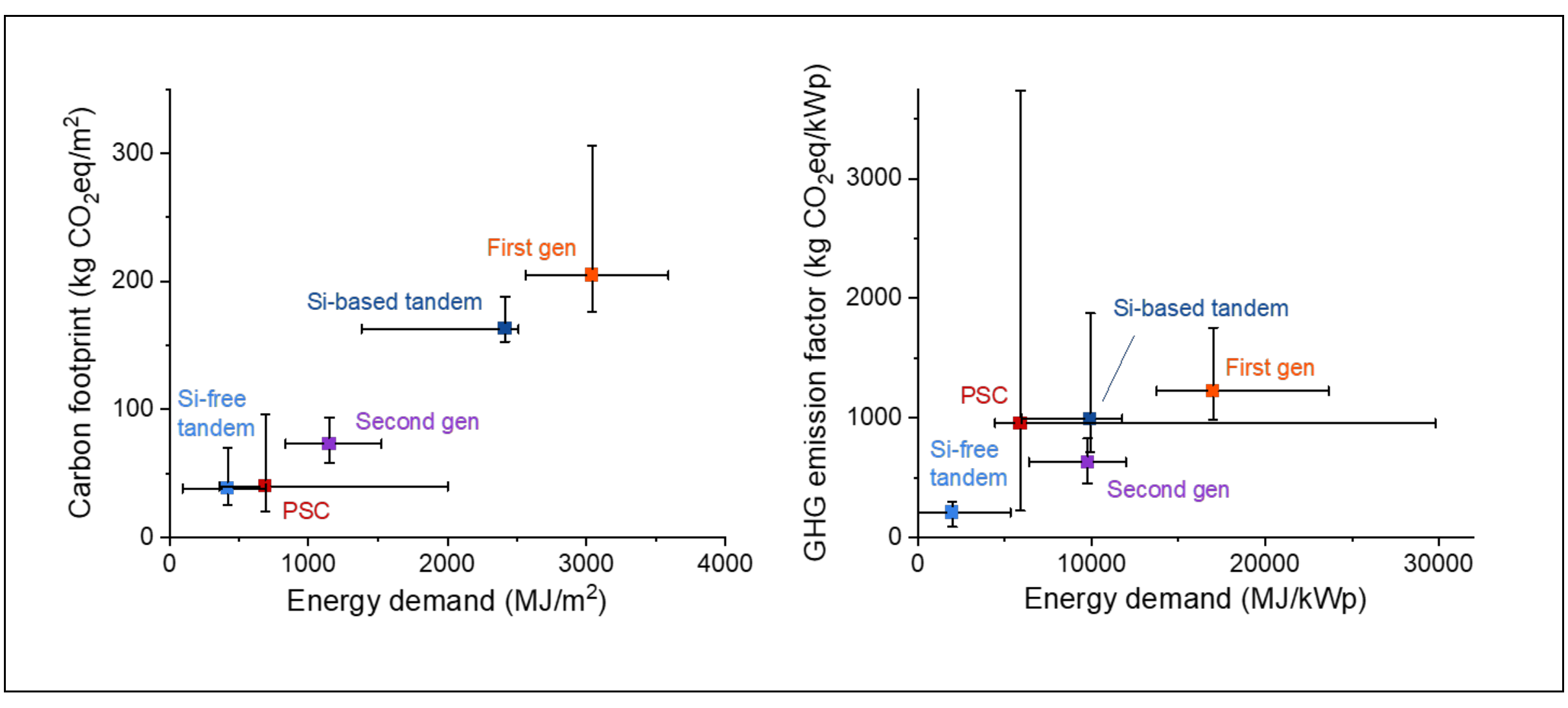

## [H1] Sustainability Challenges with MHP-Based Tandems

Whilst MHP-based tandems offer appealing opportunities to improve PV circularity and lower the overall environmental footprint per unit energy produced, it is critical to evaluate the environmental risks when tandems are deployed on the multi-$TW_p$ scale. This section discusses the sustainability challenges based on the main components of current-generation MHP/Si and all-MHP tandem architectures.

Our analysis focuses on elements that are either flagged in regulatory frameworks, *e.g.*, the European Union's Restriction of Hazardous Substances (RoHS) directive[116], Control of Substances Hazardous to Health (COSHH) or critical raw material lists (e.g., UK criticality assessment[117], EU CRM act[118], ACS Endangered Elements[119]. We calculate the mass of each metal in the selected device layers and calculate the demand–production ratio (DPR) for evaluating the supply risk (Supplementary Note 1). DPR > 20% signals a barrier to sustainable $TW_p$-scale deployment[120]. We evaluate the primary energy demand of virgin materials that provide the key elements[121], which, ignoring solvents and processing for a lower-bound estimate, suggest the essential materials energy requirement at 1 TW-scale PV deployment.

### *[H2] Transparent conducting electrodes*

Tandem solar cells require TCEs for efficient charge-carrier extraction, while allowing light to reach the underlying layers. The TCE requirement for tandems is higher than in single-junction cells because these are required at the recombination contacts for 2-T tandems, or for the top and bottom electrodes in 4-T tandems (**Box 2**). ITO currently leads the TCE market with a 55% share in 2024[122]. However, In, classified as a critical raw material, has limited global reserves and its production is energy-intensive, as it is primarily produced as a byproduct of Zn mining (**Figure 2a**)[123]. For the total materials requirements for manufacturing 1 $TW_p$ worth of solar modules, the projected DPRs of In are on average 761% for perovskite-silicon and 440% for all-perovskite tandems (**Figure 2b**). This results in supply instability and price volatility, with costs reaching ~$452/kg[122]. The high DPRs arise from the scarcity of In, and the relatively thick ITO layers required to have low sheet resistance, which elevates material demands. In our calculations, the In content of the TCO also accounts for 24–50% of the total materials primary energy demand in MHP/MHP tandems (**Figure 2c**). If MHP PVs are to be deployed at the TW-scale, it will be challenging for the PV industry to afford virgin In. Therefore, diversifying away from In in TCOs, combined with developing recovery strategies for ITO/glass substrates are urgent strategic priorities[123].

A 4-T module requires three or four thick transparent electrode layers (**Box 2**), increasing the total TCO material requirement to roughly five times that of a 2-T architecture. Many efficient monolithic 2-T tandems use one TCO as primary top electrode and ITO or indium zinc oxide (IZO) as the interconnection layer (ICL) to provide a transparent recombination contact.

A conventional 20-nm ITO ICL in a 25.4% tandem corresponds to an indium DPR of about 70%, yet sub-10-nm ITO layers often introduce resistive losses and interfacial damage that degrade device performance[124]. Amorphous IZO maintains high transparency and adequate conductivity at few-nanometre thicknesses. A 35%-efficient tandem device with a 10 nm-thick IZO-based ICL can reduce the DPR to about 23%[67], which is generally considered compatible with large-scale commercialization. The primary energy analysis of typical high-performance 2-T tandems also indicates that the In usage within the ICL does not dominate the overall resource footprint (**Figure 2d**). In contrast, alternatives involving ultrathin metallic ICL, such as Au, are far more problematic. An extremely thin Au ICL of 0.4 nm results in a DPR of approximately 1.4% and contributes up to 42% of the total primary energy demand in MHP/MHP tandem devices, making gold an unfavourable option from environmental and economical perspectives.

### *[H2] Absorber layers*

By breaking down the primary energy for the materials used in a MHP/Si tandem (Box 2), our calculations show that the single-crystalline c-Si wafer accounts for ~95% of the total the embodied energy of the tandem (**Figure 2c**), consistent with prior LCAs (Box 3)[88]. Since MHP absorbers require orders of magnitude less energy to produce than c-Si, replacing Si with a perovskite bottom cell reduces primary energy consumption by >99%, such that the extraction and deposition of the TCO becomes the most energy-intensive component in a MHP/MHP tandem.

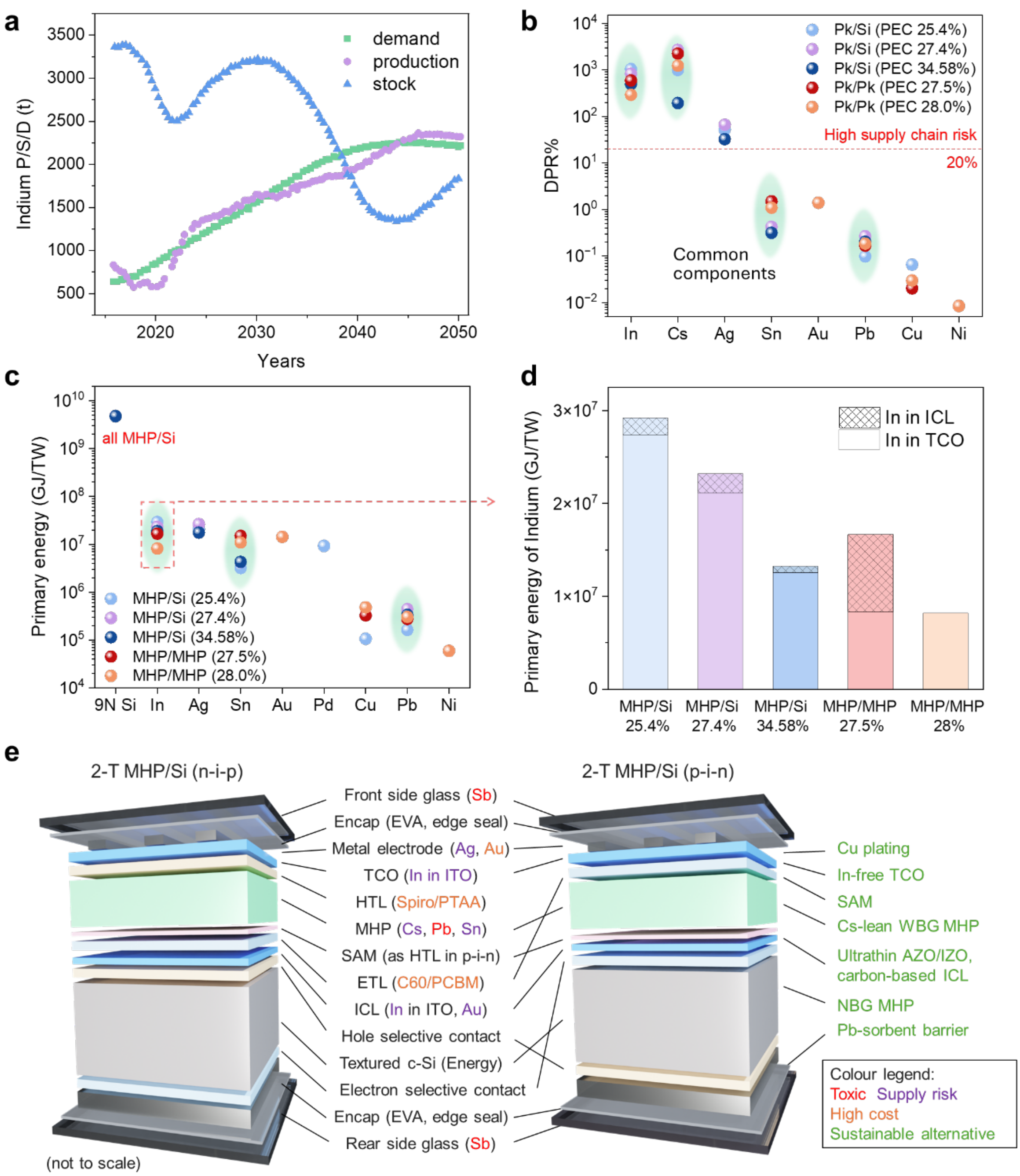

**Fig. 2 | Resource analysis of perovskite-based tandem photovoltaics**. **a,** Indium supply constraints. P refers to primary indium production, S to accumulated stock, and D to demand[123,125]. **b,** Demand-to-production ratio (comparing projected demand with current global production) and **c,** primary energy needed for virgin metal production for critical components in five high-performance tandem cell configurations (power conversion efficiencies (PCEs) labelled)[126–129]. **d,** The breakdown of indium's energy demand in the TCO and interconnection layer (ICL). **e,** Graphical representation of the two-terminal perovskite/Si tandem cell in the n-i-p and p-i-n architectures. The sustainability issues of each functional layer and their potential solutions are highlighted. Red indicates toxicity, purple indicates supply risk, orange indicates high cost, while materials indicated in green indicate a more sustainable alternative to material currently used in perovskite-silicon tandems.

Wide-bandgap (WBG) perovskites for tandem top cells typically have 10-25 mol% Cs as the A-site cation to stabilize the perovskite structure with the higher Br content, compared to 5-10 mol.% Cs required in lower-bandgap MHPs for single-junction PVs[130]. A thicker MHP top absorbing layer (approximately 1 μm) is also required to fully cover the textured c-Si bottom cell. Although Cs is a minor component by mass (2-5 wt%) in perovskite films, it is constrained at the TW-scale by an average DPR of 1468%. Whilst Cs-free WBG MHPs are available for the tandem top cell (*e.g.*, $((NH_2)_2CH)_{0.6}(CH_3NH_3)_{0.4}Pb(I_{0.6}Br_{0.4})_3$[131]), $(CH_3NH_3)_{0.96}((NH_2)_2CH)_{0.1}PbI_2Br(SCN)_{0.12}$[132]), they suffer from instability and high defect densities, necessitating further advancements in composition engineering and interface passivation[133]. Rubidium (Rb) has also been used in the A-site of the perovskite for passivation[134], but faces fundamental supply chain constraints and prohibitive costs[134]. On the other hand, Pb, which is an integral component of the absorber layer (~7–8 kt per TW), faces minimal supply risk (DPR < 0.2%), but rather faces environmental regulations, like the RoHS directive (refer to later section on preventing lead release).

***[H2] Charge transport layers***

Tandem cells incorporate several charge transport layers to facilitate the extraction of photogenerated electrons and holes from each sub-cell. HTL materials do not contain obvious critical metals in their nominal compositions (except for traces, <1 at.%, of lithium to dope $TiO_2$ ETLs), but organic HTLs such as spiro-OMeTAD, Poly[bis(4-phenyl)(2,4,6-trimethylphenyl)amine] (PTAA), and Poly(3-hexylthiophene-2,5-diyl) (P3HT) require palladium (Pd) catalysts during synthesis[41]. Pd has ample global production (DPR < 0.5%), but is known for having high embedded primary energy. Organic HTLs face issues of high cost and high environmental impact[135,136]. For example, spiro-OMeTAD from Sigma Aldrich costs US$346-482 $g^{-1}$ (price at Nov 2025), contributing to approximately 10% of the total materials

cost of a single-junction perovskite module, based on a cost per gram to cost per watt conversion using a typical HTL thickness, module efficiency (~20%), and a target module cost of ~US$0.50 $W^{-1}$ [137]. PTAA is close to ten times the cost of spiro-OMeTAD, due to its much more complex, multi-step synthesis/purification process with low yield[67]. Similarly for organic ETL materials, fullerene ($C_{60}$) and [6,6]-Phenyl-$C_{61}$-butyric acid methyl ester (PCBM) costs hundreds to more than a thousand US dollars per gram, depending on the purity[138]. There is therefore strong motivation to replace these charge transport layers with more cost-effective and more stable metal oxide HTLs (*e.g.*, $NiO_x$ and $MoO_3$) and ETLs ($TiO_2$, $SnO_2$, etc).

However, tandems have limited selection of inorganic transporting layers, because, especially for 2-T tandems, layers must be deposited using processes that are compatible with the underlying device stack. For example, c-Si bottom cells employing amorphous-Si passivation or TCO layers imposes upper thermal limit of < 200–250 °C, which limits the direct use of high-temperature $TiO_2$ or $NiO_x$ layers. Depositing on MHP bottom cells must avoid polar solvents, high-temperature annealing, and sputtering, therefore the chosen materials should be compatible with solvent-free processes *e.g.,* atomic layer deposition, evaporation, dry transfer[139].

***[H2] Self-Assembled Monolayers***

Self-assembled monolayers (SAMs) are an emerging class of ultra-thin interface materials that can improve the performance and sustainability of MHP single-junction PVs and tandems[140,141]. In a conventional n–i–p architecture, SAMs are typically used between the perovskite and the metal electrode or charge transport layer to reduce non-radiative losses at interfaces[67,142]. In inverted (p–i–n) perovskite stacks, SAMs can replace the polymer HTLs (**Figure 2e**). In the current certified world-record perovskite–silicon tandem (34.58% as of July 2025)[67], SAM serves as a hole-selective interconnection layer between the perovskite absorber and the IZO recombination contact. Furthermore, cross-linking mixed functional molecules to form covalently interconnected interfacial networks (*i.e.*, co-SAMs[143,144]) can lead to improved stability by suppressing molecular diffusion, interfacial delamination, and chemical degradation under heat and illumination.

Although no full LCA of SAM-based contacts has yet been published, they are widely considered more sustainable than using organic charge transport layers because they use orders of magnitude less material, *i.e.,* sub-nanometre SAMs versus conventional >10-nm-thick polymeric PTAA or doped Spiro-OMeTAD HTLs. SAMs also offer simpler synthesis than polymeric materials. [2-(9H-carbazol-9-yl)ethyl] phosphonic acid (2PACz) is one of the simplest and most scalable SAMs, synthesized in three steps from inexpensive precursors[142]. Because SAMs spontaneously form ordered monolayers upon solution deposition, they eliminate the need for hygroscopic dopants required in organic HTLs, thereby reducing compositional variability. Using conjugated halogen-functionalized SAMs can achieve uniform coverage on rough pyramidal surfaces, achieving improved hole extraction and effective defect passivation on textured interfaces[145]. Although most high-efficiency tandems remain limited to small areas, SAM deposition is inherently compatible with scalable techniques such as dip-coating, slot-die coating, spray coating, and vapor-phase self-assembly[146]. Moreover, surface pretreatments required to promote SAM attachment involve minimal energy input compared with the high-temperature or vacuum-based processes, thereby preserving the low thermal-budget advantage.

***[H2] Cell-to-Module***

Going from individual cells to integrated modules, Ag is commonly used as a conductive paste for the front and rear electrodes, as well as for connections between modules. Although Ag is not currently classified as critical by the EU and UK, it is of strategic importance because the PV industry consumed 16% of the global Ag supply in 2023[147]. The trend in the PV industry towards technologies like TOPCon (12.26 mg $W^{-1}$ Ag) and heterojunction cells (16.7 mg $W^{-1}$ Ag)[148] further raises the baseline silver consumption with DPR of 46% and 63% for a one $TW_p$-scale PV industry, claiming half of the world's entire annual mined silver supply. However, this industrial-scale metallization method (screen-printing silver pastes) is not compatible with the low-temperature requirements of the perovskite layer. In lab-based MHP cells, silver consumption is mostly determined by the electrode layer thickness (100–300 nm) and coverage, estimated to have a DPR of 32-66% if not counting the substantial materials wastage (sometimes over 90%) during thermal evaporation. Au electrodes are common in record-performance MHPs but unfeasible at large scale due to being approximately 80 times more expensive than silver and posing severe environmental risks from mining, where ore extraction and processing release toxic metals, such as arsenic, mercury, and zinc into soil and water systems[149]. Promisingly, tandem cells possess a lower operating current that significantly reduces resistive power losses, making them more tolerant of higher-resistance grids, and encouraging silver-lean PV technologies, such as ultra-fine-line screen printing, silver-lean intermittent contact patterns, and substitution of silver with Cu[150].

As the protective cover in PV modules, tempered low-iron glass (~2–3 mm) often employs antimony trioxide ($Sb_2O_3$) to improve optical clarity. $Sb_2O_3$ is toxic, carcinogenic, and complicates recycling, prompting calls for Sb-free formulations and safer end-of-life handling[151]. Furthermore, Al, mostly used to frame the PV module, has its naturally occurring ore (bauxite and feldspar) listed critical in the EU Critical Raw Materials Act[63]. While not currently posing a supply-chain risk, there are calls for good practice in recycling the Al from used Si PVs. 4-T configuration modules require an additional layer of structural glass that further increases the module mass and environmental footprint. Overall, **Figure 2e** summarizes the key sustainability challenges in the constituent functional layers in tandems, factoring in raw materials criticality (In, Cs, Au and Ag), primary energy demand (Au), cost intensity (organic HTLs and Au), and toxicity (Pb and Sb from $Sb_2O_3$). The potential pathways and material substitutions that could alleviate these challenges are indicated.

## [H1] Materials Strategies to Enhance Sustainability

In this section, we examine strategies to enhance the sustainability of perovskite-based tandem solar cells, focusing on three critical challenges: the development of In-free transparent electrodes, the recycling and reuse of materials, and prevention of lead release during PV deployment. These areas represent the most pressing sustainability concerns, as they directly impact materials scarcity, environmental footprint, and long-term device safety, while also influencing the commercial viability of high-efficiency tandem technologies.

### *[H2] In-Free Transparent Conducting Electrodes*

Motivated by the need for more sustainable and scalable alternatives to ITO in tandem photovoltaics, we here focus on In-free transparent conducting electrode (TCE) materials, including oxides (TCOs) and non-oxides. We discuss different classes of In-free TCEs and the fundamental factors that determine their performance and limitations, with particular attention to their compatibility with MHP/c-Si tandem architectures.

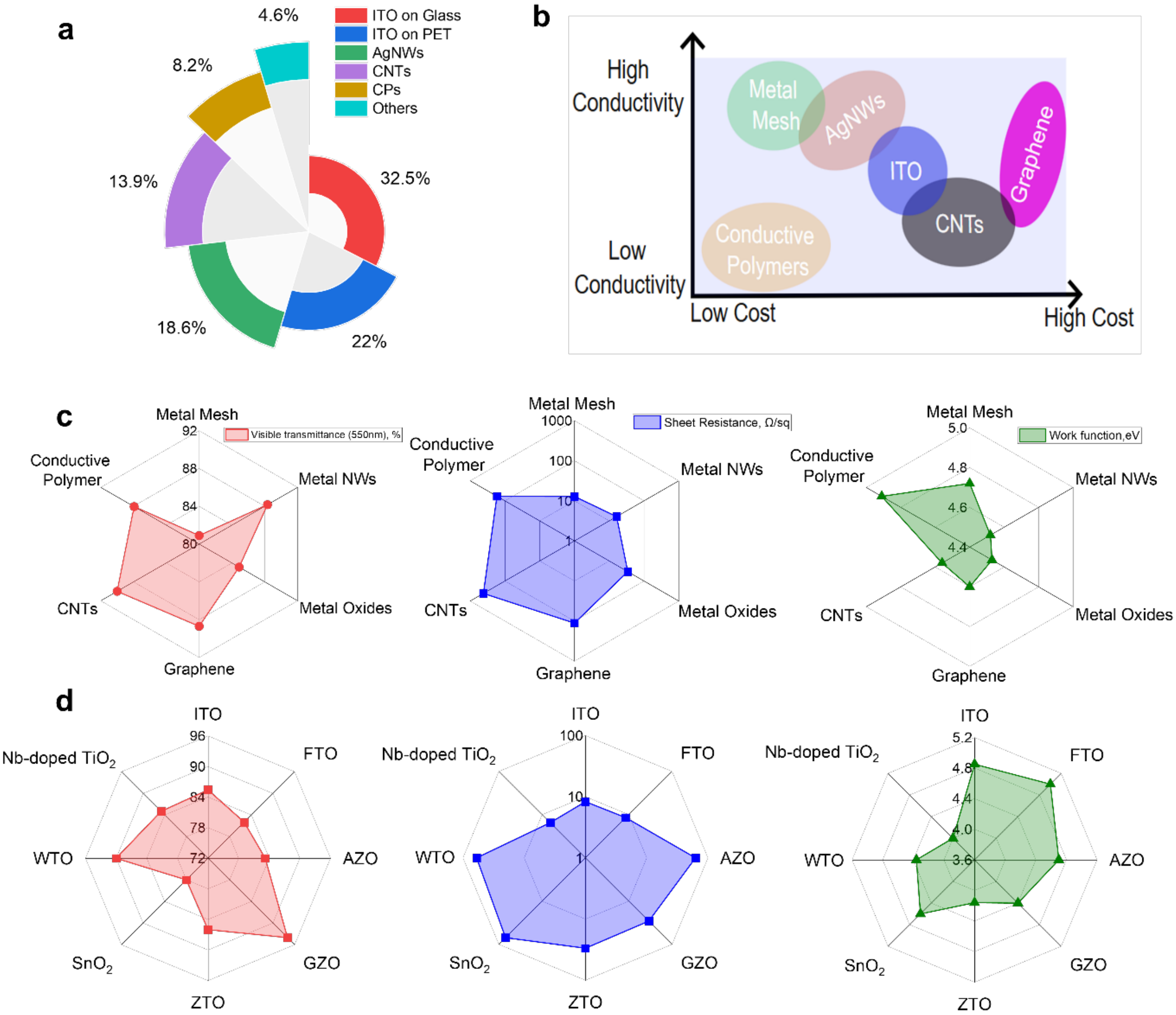

**Fig. 3 | Overview of the TCE landscape for tandem solar cells. a,** Market distribution of major TCE materials, illustrating the relative commercial share of ITO (on glass and PET), Ag nanowires, carbon nanotubes (CNTs), conducting polymers (CPs), and other emerging alternatives[122]. **b,** Comparison of major classes of TCE materials based on their cost and conductivity. Data shown in Supplementary Table 2. **c,** Radar diagrams comparing the performance of transparent conducting materials (TCMs) relative to ITO across key criteria, including conductivity, cost, transmittance, stability, scalability, and compatibility with perovskite processing. The detailed data used for these radar diagrams are provided in Supplementary Note 3.

In-free TCEs for MHP tandem solar cells have garnered growing interest as the field progresses toward commercialization and record-breaking efficiencies. ITO remains the benchmark TCE due to its high visible transmittance (85-90%)[152,153], and electron mobility of 27-50 $cm^2 V^{-1} s^{-1}$ [154,155], although related indium-based oxides such as indium zinc oxide (IZO) and indium gallium zinc oxide (IGZO) are also used in tandem architectures[156–159]. Yet, the reliance of these TCEs on scarce indium with high DPR motivates the development of more sustainable alternatives (**Figure 3a**)[122,160], with silver nanowires (AgNWs), carbon nanotubes (CNTs), and conducting polymers (CPs) capturing 18.6%, 13.9%, and 8.2% of the global transparent conducting films market in 2024, respectively[122]. Emerging materials, such as graphene, CNTs, and CPs also occupy distinct regimes in the cost-conductivity relationship compared to In-based TCEs (**Figure 3b**)[161–165]. For instance, graphene (~89% visible transmittance, ~113 Ω $sq^{-1}$ sheet resistance)[166,167], CNT networks (~90%, ~417 Ω $sq^{-1}$)[168,169], and conducting polymers (~88%, ~170 Ω $sq^{-1}$)[170–172] differ from ITO (~85% transmittance, ~8 Ω $sq^{-1}$)[173], as shown in **Figure 3c**. MNWs offer excellent flexibility and high conductivity (~88% transmittance, ~16 Ω $sq^{-1}$)[83,161,174,175], CNT networks provide mechanical resilience and chemical stability, and CPs enable low-temperature solution processing; however, no single alternative yet combines the optical transparency, low resistivity, environmental stability, surface uniformity, and industrial maturity of ITO[176–178]. In tandem devices, electrode deposition must also avoid damaging the perovskite absorber or disrupting the temperature-sensitive passivation layers used in c-Si bottom cells.

Fluorine-doped tin oxide (FTO) has also been widely used as a TCO due to its thermal and chemical robustness, particularly in silicon bottom cells and tandem device architectures[179]. However, $SnO_2$-based TCOs generally exhibit lower intrinsic mobility, requiring higher carrier concentrations (≥$10^{20}$-$10^{21}$ $cm^{-3}$)[180,181] to achieve low resistivity, which enhances free-carrier absorption and shifts the plasma edge into the near-infrared, reducing NIR transmission relative to higher-mobility TCOs[182]. Although alternative tin-based oxides such as tungsten-doped $SnO_2$ (WTO) can exhibit high optical transparency (~85–90%) and sheet resistances typically in the tens of Ω $sq^{-1}$ [183,184], their implementation in tandem devices remains limited. Additionally, high deposition temperatures (>400 °C)[185] can degrade perovskite layers and charge-transport materials, limiting the applicability of FTO-type electrodes in monolithic tandem architectures, particularly as top contacts. This is particularly challenging for tandems as the high deposition temperatures can degrade the perovskite absorber and compromise the amorphous silicon passivation layers of the cells.

Zinc oxide–based TCOs, particularly aluminium-doped zinc oxide (AZO) and gallium-doped zinc oxide (GZO), are promising In-free alternatives. Substitution of $Zn^{2+}$ with $Al^{3+}$ or $Ga^{3+}$ introduces n-type carriers, while the low effective electron mass ($0.24\text{-}0.3m_0$)[186] in ZnO supports relatively high mobility and adequate conductivity at lower carrier concentrations, improving near-infrared transparency for optimized GZO films[187]. These materials can achieve sheet resistances of 29–62.5 $\Omega$ $sq^{-1}$ and visible transmittance of approximately 83-94%[188–192]. Zinc tin oxide (ZTO) has also been explored as a related indium-free TCO, exhibiting ~86% visible transmittance[193] and sheet resistance around ~30 $\Omega$ $sq^{-1}$ [194], although its conductivity generally remains below that of optimized AZO or GZO films. Perovskite-silicon tandem cells with AZO-based TCEs have demonstrated PCEs exceeding 28%, close to those achieved using ITO[188]. Nevertheless, lower carrier mobility (15-30 $cm^2$ $V^{-1}$ $s^{-1}$) compared to ITO limits the ability to reduce the resistivity (**Figure 3b**)[195,196], and challenges remain in depositing these films without damaging underlying perovskite layers, managing interface defects, and ensuring long-term thermal and moisture stability[187]. These materials are generally more compatible with tandem processing because they can be deposited at lower temperatures, although sputter damage to MHP layers and amorphous Si passivation layers for c-Si remains a challenge.

Although no single material currently surpasses ITO across all requirements for TCEs (**Figure 3d**), progress in hybrid electrodes (e.g., metal-nanowire/oxide and carbon-based contacts), low-temperature printable TCEs, and emerging ITO/indium recycling routes are beginning to broaden the sustainable TCE landscape. These innovations will enable tandem solar cells to achieve high efficiency and environmental sustainability. Another promising route to achieve high performance, stability and sustainability is to develop hybrid electrodes, such as AgNWs embedded within metal oxides or CPs[161,197]. These hybrid electrodes may facilitate tandem integration by enabling lower-temperature and solution-processed electrode fabrication. The optimal AgNW density for replacing TCOs has been estimated to be approximately 120 mg $m^{-2}$, balancing high optical transmittance with low sheet resistance[198]. Assuming a PCE of 25%, this Ag loading corresponds to a DPR of 2.1% (calculation detailed in Supplementary Note 2), rendering AgNW networks a highly feasible alternative to alleviate In usage. Indeed, state-of-the-art In-free tandem architectures are already approaching this 25% threshold; for instance, perovskite/silicon tandems utilizing ALD-deposited AZO as an interconnection layer have recently demonstrated sub-cell efficiencies of 24.94%[199]. While this remains below the record 35.0% achieved with optimized ITO/IZO stacks[200], the ~10% relative efficiency trade-off is increasingly offset by a 90% reduction in material criticality. The environmental footprint of these hybrids, however, depends heavily on their synthesis routes[201]. Photonic curing, which uses millisecond light pulses for rapid, low-energy processing, enables scalable, roll-to-roll fabrication on heat-sensitive substrates. In parallel, work function engineering through doping or interfacial buffer layers improves device efficiency and long-term stability.

## *[H2] Preventing lead-release*

A critical factor to address is the inherent use of neurotoxic Pb in lead-halide perovskites. Given the water solubility of MHPs, $Pb^{2+}$ can readily be released to the environment upon real-world deployment through contact with water, or released in the vapour phase during a fire. However, there remains debate about how severe the release of Pb would be. Whilst some studies have indicated that plants grown in soil contaminated with 250 mg $kg^{-1}$ $Pb^{2+}$ from MHP films exhibits orders of magnitude increases in their Pb content[202], other works indicate that $Pb^{2+}$ should be immobilized in soil[203]. Furthermore, given that thin films are used, the content of Pb in MHP thin films is approximately an order of magnitude lower than that already present in c-Si solar cell modules, which use Pb/Sn solder[204]. The company Singfilm estimates that the Pb content of their MHP single-junction PV modules is 30 ppm, while Saule Technologies reported a Pb content on the order of 1 ppm[205]. These PV modules would therefore meet the requirements of the RoHS directive (1000 ppm Pb limit)[206]. However, these low calculated Pb concentration values were likely obtained by factoring in the substrate and encapsulation used in the device into the calculation. When it comes to safe MHP deployment in single-junction PVs or tandems, a key factor is how much the perovskite cell would contaminate fresh water supplies. Pb release from the aqueous or vapour phase can be suppressed through physical barriers, such as glass encapsulation. But these physical barriers can be broken during deployment, for example through hail. Strategies to prevent Pb release in the event of a catastrophic failure of encapsulation are essential, and Pb sequestration is a key approach to doing this.

Pb sequestration materials typically have functional groups that bind to $Pb^{2+}$, and include polymers, organic molecules (*e.g.*, porphorins)[207], metal-organic frameworks (MOFs), and phosphonic acids and salts (*e.g.*, hydroxyapatite)[208]. These materials are applied within the bulk of the MHPs, at interfaces, on top of the device stack, or as a coating within or outside the encapsulation[204,209]. An example of the latter is pyramidally-textured phosphate-buffered functionalized polymer film (PFPF), which can be applied to the back of the glass substrate in a MHP PV device. The glass substrate was broken by applying a 45 mm diameter ball (to simulate hail stones), and the device soaked in water. By applying the PFPF, the content of $Pb^{2+}$ leached into water decreased from 12.39 mg $L^{-1}$ to 10.1 mg $L^{-1}$, within the US requirements of <15 mg $L^{-1}$ for safe drinking water[210]. Other similar approaches include applying cation-exchange resins on top of metal electrodes[211] (with sulfonate functional groups to immobilize $Pb^{2+}$). However, the limitation of applying the Pb sequestration material on top of the device stack is that it could be physically damaged, or have the chemical sites poisoned through external contaminants, such that they are inefficient for Pb sequestration.

Alternative approaches include making the charge transport layers capable of sequestering Pb. For example, alkoxy-poly tetra ethylene glycol was developed as a hole transport layer that could capture >80% of $Pb^{2+}$. Nanoporous $TiO_2$ sponges have also been developed[212]. The Pb sequestration material could also be applied within the bulk of the perovskite. For example, acrylamide additives have been added to the bulk of the MHP films to form a polymer network, with carbonyl groups to bind to $Pb^{2+}$ [213]. Similarly, a cross-linking supramolecular complex was incorporated into the MHP film, and the high density of hydroxyl and carboxyl groups in the supramolecular complex enabled $Pb^{2+}$ sequestration[214], decreasing the Pb leaching rate in flowing water by a factor of 18. As a result, the Pb concentration in water reached 14 ppb after 1 h (within the 15 ppb limit for drinking water by the US Environmental Protection Agency), even if the perovskite film were scratched[214]. However, an important question with all of these approaches is how long these sequestration materials are effective for before all functional sites are saturated. Longer-term studies are therefore important to evaluate and guide the future development of effective Pb sequestration materials.

***[H2] Recycling and Reuse***

There are many routes being developed for recycling MHP-based devices, and the specific details of the chemistry behind these processes are covered in earlier reviews[28]. Here, we take a higher-level view of the technological, economic and policy factors that will need to be considered to realize circularity with MHP-based tandems commercially.

As discussed in earlier sections, the operational lifetime of MHP/Si tandems is not currently well-established, and lifetime impacts sustainability and economic metrics whilst also influencing when there will be significant end of life modules that require recycling or repurposing. The lower stability of MHPs compared with c-Si has led some researchers to investigate methods of removing the MHP cell from a degraded MHP/Si tandem to bring the silicon cell back to its original performance[106] and allow a new MHP cell to be deposited, regenerating the tandem. Whilst this can work for lab-based cells, the economics of this process need to be further explored, particularly regarding transport, local labour rates, impact of low production volumes and variabilities in module design. The lower capital costs of MHP processing equipment[215] (compared to c-Si) may facilitate local refurbishment of MHP tandems, although further work is needed to ensure recycled modules match the highest-performance modules manufactured in specialist facilities.

So far, despite many lab-based processes showing promise for fully circular MHP utilization[27], many engineering challenges remain: **i.** Removing the aluminium frames in a cost-effective manner, which usually takes place before delamination, can cause the chemically toughened glass to crack, reducing any opportunity for reuse in a panel. Delamination of the module, which is a common challenge for Si modules and MHP/Si modules, can further complicate reuse if the glass or cells are broken[35]. **ii.** The TCEs will be patterned in a manner specific to the module configuration; this will be specific to the module manufacture and likely to change overtime. Therefore, for the TCE to be valuable for future reuse, methods of repairing / adjusting the conducting pattern on reused glass would need to be developed. **iii.** There are many different MHP architectures, containing a wide range of materials in the ETL layers, active layers, HTL layers and top contacts[216]. Identifying the materials ahead of recycling requires a full breakdown available from the manufacturer. Legislation could enable this by requiring a breakdown of materials used in PV panels to be available. **iv.** c-Si panels, when decommissioned for repowering, face significant barriers to entering second-life markets[217], primarily driven by a 'trust gap' regarding panel performance. The invisible defects of the reused panels will lower module efficiencies and production yield, reducing the overall cost benefit. Since the c-Si module is a small part of the full system cost, manufacturers will prefer to have new panels unless incentivized through local policy measures. Reused or refurbished MHP/Si modules are likely to encounter similar issues, or face greater challenges because high-performance tandems require complete removal of the MHP and organic residue from the c-Si bottom cell.

Globally, there is variation in PV EOL regulations[115]. Whilst WEEE[218] policy in Europe dictates that c-Si PV should be recycled, there is a shortage of recycling facilities that can extract the full value from silicon panels cost effectively. Within the EU[219], the definition of 'recycling' and subsequent obligations vary, often resulting in products downgraded to construction filler materials ('downcycling')[220]. In the USA, generalized regulation comes under federal Resource Conservation and Recovery Act (RCRA) but regulation specific to PV waste varies between states, such as the SB 489 Solar Panel Recycling Program (California)[10]. In Australia, PV waste is regulated by the 2020 Recycling and Waste Reduction Act[221]. In China there is no obligation to recycle solar panels, but guidance is covered by the solid electrical waste directive GB/T 23,685-2009 and GB/T 38785-2020 for building integrated panels, but this does not mandate recycling[222]. In 2021, South Africa introduced EPR regulations for PV[223]. This limitation is mainly due to the current small volumes of solar panels making centralised recycling facilities unprofitable, such that recycling is not cost effective compared with the value of materials recovered[224,225]. This is mirrored in the US where the recycling cost is ~$25 per module, and yet yields materials with a value of only ~$3, depending on recycling efficiency[207].

In addition to technical and economic challenges, there can be societal hurdles to module recycling and reuse. Extended warranties can encourage consumers to choose reused modules, however there are often secondary concerns, such as aesthetics which influence consumer behaviour[207]. Furthermore, the solar reuse market is competing with the decreasing costs of newer modules with increasing generation outputs[226]. Without enforced regulation and globally recognized standards of reused modules, there is a risk that unsafe or damaged modules which have reached EoL will be transported to developing regions without adequate local recycling facilities. Lack of local manufacturing and/or certification facilities can mean that the transport of panels is prohibitively long, reducing any cost or sustainability benefits. In conclusion, work undertaken to develop solution-based methods for recovering perovskite materials has shown there is potential to produce high quality recycled materials (see earlier section titled *Recycling Strategies*). However, the engineering issues that currently trouble c-Si PV about differences in panel composition, low material value and difficulties in delamination will be amplified with MHP/Si devices (**Figure 4**). There are significant technical challenges, and some are contradictory. For example, improving encapsulation to prevent Pb release to the waste stream can also make delamination of glass more difficult. Whilst panels are 'recycled' in Europe, often glass is down cycled as filler materials. Different glass markets (*e.g.*, China versus UK) have different composition requirements for glass, which can make recycling in a country with different regulations to the country of manufacture difficult. At present the technical challenges add a significant cost to the recycling method, which is difficult to recoup from the raw materials present in the panel, particularly with the reducing material content per $W_p$ of high-cost material like silver. Fragmented global recycling policies make it difficult to design for EOL or reuse of panels. There must be a significant change in the global policy frameworks in order to support a future scenario which has design for disassembly and profitable and responsible end of life policies at its heart.

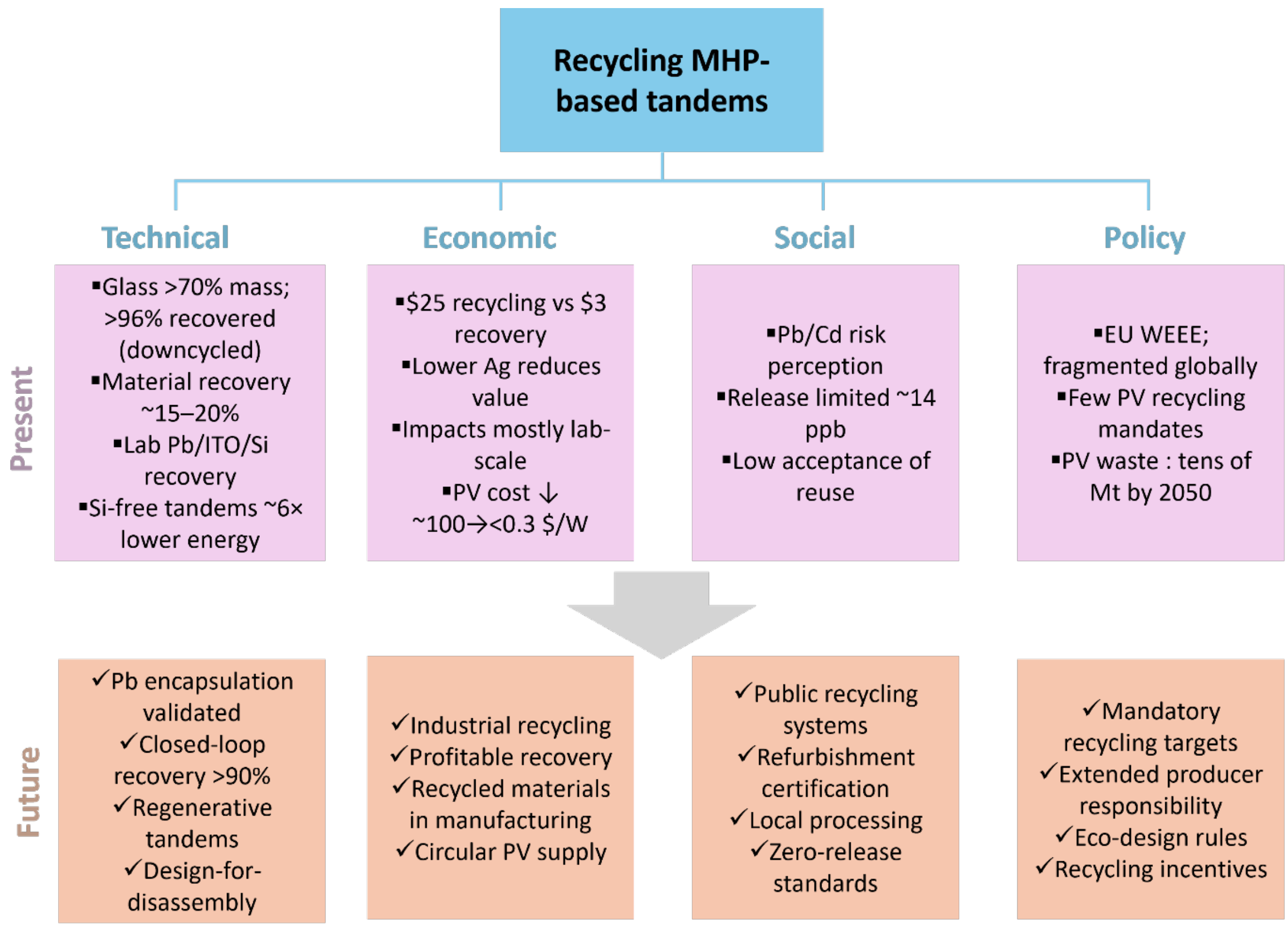

**Fig. 4 | Recycling metal-halide perovskite (MHP)-based tandem solar panels.** Current recycling is constrained by limited recovery of high-value materials, economic imbalance, public concern over lead, and fragmented regulation. Future circular deployment will require closed-loop material recovery, regenerative device design, economically viable recycling, social acceptance of reuse, and coordinated policy frameworks including extended producer responsibility and eco-design.

## [H1] Future Policy Challenges

With the expected TW-scale deployment of MHP-based tandem devices over the next decade, we recommend coupling technological advances with a more nuanced and appropriate regulatory framework in which they operate. The policy framework should reflect the global nature of the solar market, whereby solar manufacturers operating worldwide standardize their products to meet the most stringent regulations, making them universally compliant[227]. Foreseen policy challenges and the financial mechanisms that they operate in should be overcome together with the advancement of the technology itself. There is an urgency to pre-empt developing effective policy before MHP-based tandems are widely deployed, recognizing there is typically a delay between rapidly emerging technologies entering the market and regulations entering into legislation and adoption. Stakeholders and policymakers have the benefit of hindsight and learnings from decades of growth from the c-Si, CIGS and CdTe markets to acknowledge mistakes and improve the policy landscape for perovskites. For example, in the UK, the WEEE regulations came into force in 2014, but government incentives for PVs were introduced in 2010, meaning there were four years of exponential PV deployment in which the financing or reporting of waste volumes were unaccounted for.

Looking forward to the future deployment of MHP-based tandems, the first question that needs to be addressed is which policy framework will regulate the MHP sector. Questions have already been raised by the solar sector as to the appropriateness of PV being included in the WEEE regulations, due to stark differences between electronic consumer products and solar PV technologies. For example, PVs have longer lifespans, electricity is generated rather than consumed and are directly impacted by global events and energy markets[228]. The existing EPR and WEEE regulations do not account for variations in solar PV cell types (e.g., CIGS, CdTe, c-Si), which are viewed ubiquitously in the policy framework, despite diverging characteristics, degradation modes, and EOL pathways[229]. The integration of more complex material compositions, MHP cell structures, and more durable encapsulants will require a new approach as to how perovskites are regulated. For example, it may be necessary for MHPs to be kept separate from traditional c-Si waste streams, due to the need to isolate the reusable elements from EOL for producing new panels, and this would require separate infrastructure, collection streams, and finance mechanisms compared to what currently exists for c-Si PV panel recycling.

Within the European level, there are variations between how waste is financed and how the WEEE Regulations are implemented[230]. For example, in Belgium and France, Pay as You Go (PAYG) ('visible fees') apply for WEEE products, but in the Netherlands, UK, Germany, or Italy a 'Pay as you Throw' (PAYT) model is adopted[220]. In other finance models, First Solar, which manufactures CdTe PVs, provides and manages their own takeback and recycling scheme for their products, which is incorporated into the final price for the consumer. However, there is a risk with the 'First Solar' model if the company supplying the future MHP-based tandem modules runs into administration or mishandles their EOL responsibilities or finances, it is appropriate that an inter-governmental watchdog monitors or has oversight over this type of scheme. The perovskite industry needs to understand and engage in the discussion over the appropriateness of the regulations that will govern it, and the finance model which will fund its EPR.

The degradation mechanisms of MHPs are not yet well understood, and this uncertainty will impact on policy. Organizations, such as the International Electrotechnical Commission (IEC) and American Society for Testing and Materials (ASTM) International, are addressing the need for MHP-specific standards, which include essential characteristics, such as encapsulation durability, stability under harsh circumstances, and energy output lifetime[231]. The existing IEC standards for stability tests of PV devices are designed for conventional solar panels based on c-Si, GaAs, CIGS, or CdTe. However, MHPs have different device architectures and specific material properties and therefore, a new standard stability test should be established for MHPs under illumination, heat, and bias voltage[232]. It is a recommendation of this paper that MHP manufacturers should provide mandatory Environmental Product Declarations (EPD). EPDs provide transparent, third-party verified documentation in a standardised format, that can include information such as LCAs, sourcing of materials (*e.g.*, use of scarce elements), embodied carbon / GWP, environmental impacts, and energy intensity of manufacturing processes, or new processing sequences required for EOL recycling. Mandatory EPDs would provide greater transparency for the MHP industry to help buyers make informed decisions about the products' environmental credentials. If this approach were enforced, this would create a level playing field for all manufacturers to design for sustainability from the outset. Manufacturers would be incentivized to innovate perovskite technologies with decreasing impact on the environment, instead of the current status quo of competing on ever-increasing cell efficiency and energy yield.

Globally, there are additional regulatory frameworks and initiatives that may intersect with the growth of perovskite technology, including RoHS, the CRM Act, ACS Endangered Element Act (ACS EEA), and the Eco-design for Sustainable Products Regulation (ESPR), Digital Product Passports, occupational health *e.g.*, Control of Substances Hazardous to Health (CoSHH), and Solar Stewardship Initiative (SSI). Some of these initiatives, primarily developed for the c-Si market, may inadvertently be the panacea that perovskite policy challenges need around sustainability challenges discussed in this paper. However, the development of a responsible and coherent policy framework is most effectively implemented by the perovskite sector itself via voluntary standards with self-preserving interests. However, by virtue of being voluntary does not guarantee compliance. Whilst the benefits of MHPs offer advantages for rooftop applications with higher performance modules, circularity means observing the waste hierarchy, where the first responsibility is prevent. In this respect, the perovskite market should be vigilant of its potential role in encouraging early replacement of modules and should work in coordination with c-Si takeback schemes for responsible reuse and recycling of older c-Si modules.

## [H1] Conclusions and Outlook

MHP-based tandem photovoltaics present a promising pathway to combine high efficiency with approaches that enable low environmental impact, and circularity in manufacturing. Their intrinsic advantages include low-temperature processing, simple device architectures, and strong prospects for material recovery. However, sustainable TW-scale deployment requires coordinated advances across the materials used, manufacturing practices, improving the durability of the devices, and end-of-life management, together with an effective international policy framework for incentivizing circularity. Recent progress shows that circularity and performance can advance in parallel, with developments in green solvent systems (*e.g.*, water-based routes) for Pb recycling, effective sequestration strategies to prevent Pb release, and regenerative tandem designs that allow MHP top-cell replacement. Stability targets, durability qualification, and recycling infrastructures must evolve alongside efficiency improvements to avoid repeating the waste, resource, and design constraints faced by c-Si PV. In addition, policy frameworks developed for legacy PV technologies must adapt to the distinct material flows and risks of MHP tandems, incentivizing eco-design and full material traceability. With strategic alignment between research, industry, and regulation, MHP-based tandems can support the development of sustainable photovoltaic technologies at terawatt scale.

When considering priorities for future tandem photovoltaics, materials selection emerges as a central sustainability challenge, particularly due to the reliance on elements with high depletion potential or supply risk, such as In, Ag, Au, and Cs. In, widely used in TCOs, faces both geological scarcity and competing demands from display technologies, motivating the development of In-free transparent conductors. Promising alternatives include doped wide-bandgap oxides (e.g., Al- or Ga-doped ZnO, and W-doped $SnO_2$-based TCOs), ultrathin metal meshes, conductive polymers, and carbon-based electrodes. Yet challenges persist in these alternatives, including stability under illumination and heat, interfacial compatibility with MHP and silicon, and scalable, low-temperature processing. Similarly, replacing Ag and Au grids with copper-based contacts or printed carbon electrodes can significantly reduce the environmental impact of the tandem PV module, but requires advances in corrosion resistance, contact resistivity, and long-term reliability. Looking ahead, developing high-performance indium-free TCOs offers a path to combine sustainability with efficiency in tandem solar cells. This includes adapting doped ZnO and $SnO_2$ thin films via mild, tunable processes (*e.g.*, conventional/spatial atomic layer deposition[161,233–235]) exploring new dopants, and investigating alternative families of materials like conductive polymers, which offer additional advantages for flexible or lightweight devices. Cost, stability, and scalability remain key considerations in all approaches.

For the solar absorber, future efforts should focus on developing high-capacity sequestration strategies that can readily be integrated into modules with scalable, low-temperature, low-solvent approaches, and validated under realistic mechanical, environmental, and cycling conditions, to ensure negligible lead release over the lifetime of the MHP-based tandem module. When considering EOL, MHP/Si tandems overlap with c-Si PV in some of the challenges. That is, design for longevity can make the module more difficult to recycle, whilst technological developments (such as increased efficiency) can lead to 'repowering' and early retirement of modules after less than 10 years operation. Although there is a potential market for second life PV panels, there are many barriers preventing reuse, such as warranties, transport and certification. These barriers are not constant over time and supply restrictions of new panels could change the outlook for second life panels in future. Methods to reduce the environmental impact of producing MHP-based tandems can be both positive and negative for recycling at EOL. For example, the development of materials that can be dissolved in lower toxicity solvents is an advantage. Reducing the Ag content per panel is important for enabling TW-scale deployment, however, lower Ag content also reduces the economic incentives for recycling panels, as Ag is one of the main high-value materials recovered during PV recycling. For economic recycling, and ease of separation of different components are key. This covers not just the module itself, but also the framing, junction boxes and any integrated connections and diodes.

As such, we strongly recommend taking a systems approach to realize the optimal environmental performance of MHP-based tandems. The central message is simple: sustainability does not have to trail performance. By treating recyclability, critical-material substitution, and lead risk management as first-order design constraints, alongside the more widely-studied efficiency and durability metrics, the MHP tandem community can turn circularity from a compliance obligation into a competitive advantage. Designing the devices, standards, and takeback systems together could make future commercial MHP-based tandems green not just in their electricity production, but across their entire lifecycle.

## Acknowledgements

A. S. and R. L. Z. H. thank support from St. John's College Oxford for support through the visiting researcher scheme. The authors acknowledge the Agence Nationale de la Recherche (ANR, France) for support via the project ANR-23-715 CE51-0029-01 "INNOVATION". A.S. would like to thank Toulouse INP for the ETI POLYTANCE project and the CNRS for the PROCEDO project. All authors are grateful to NREL team for providing the base data set for the module toxicity and waster review. R.S.B was supported by the Leverhulme Prize for Engineering (PLP-2022-154) and the UK Engineering and Physical Sciences Research Council grant number EP/X037169/1. S. D. and R. L. Z. H. thank First Solar for financial support. J. B. acknowledges funding from the UKRI EPSRC grant fund TReFCo EP/W019167/1. M.B acknowledges PhD studentship funding from ReSolar® Limited and T.M acknowledges funding from UKRI EPSRC funded grant (EP/X038823/2). R.L.Z.H. thanks the UK Research and Innovation for a Frontier Grant (no. EP/X029900/1), awarded via the 2021 ERC Starting Grant scheme. He also thanks the Royal Academy of Engineering and Science & Technology Facilities Council for support through the Senior Research Fellowship scheme (no. RCSRF2324-18-68).

**Author contributions**

R.L.Z.H. and A.S. conceived of the idea for this review. J.B. co-wrote (with M.B. and S.D.) the sections on recycling and policy, and contributed to editing and review. MB wrote the sections on *Sustainability Challenges in the Photovoltaics Industry* and *Future Policy Challenges*. S.D. wrote the section on *Sustainability Challenges with MHP-Based Tandems*. S.B. wrote Box 2 and the section on opportunities with MHP-based tandems for sustainability, with contributions from S.D. A.S. wrote the section on In-free TCEs, while R.L.Z.H. wrote the section of Pb sequestration. R.L.Z.H., A.S., J.B., M.B., S.D. and T.M. all contributed to the discussion of the content, and worked on refining the writing of this review together.

**Competing interests**

R.L.Z.H. is CTO of *NanoPrint Innovations Ltd.*, a startup developing spatial ALD reactors for manufacturing thin films used in photovoltaics. Aside from this, the authors declare no other competing interests

**Supplementary Information**

Details of the calculations of DPR and performance of TCMs across their key metrics

**Supplementary Data**

Comparative summary of life-cycle energy and environmental performance indicators for different photovoltaic (PV) technologies, compiled from multiple literature sources. The table reports PV type, key structural or processing characteristics, power conversion efficiency (PCE), system boundary definitions, assumed insolation, performance ratio (PR), electricity mix, active area, system lifetime, processing energy consumption, area- and capacity-normalized energy demand, carbon footprint, energy payback time (EPBT), embodied energy, and global warming potential (GWP). Values highlighted in purple are calculated in this work from the reported values and system performance parameters, whereas all other values are directly taken from the original literature sources. Differences among studies in system boundaries, geographical context, and methodological assumptions should be considered when interpreting the results.